\documentclass[11pt]{article}

\usepackage[margin=1in]{geometry}

\usepackage{setspace}

\usepackage{hyperref}

\usepackage{amssymb}
\usepackage{amsmath}

\usepackage[hang,flushmargin]{footmisc} 

\usepackage{dcolumn}
\usepackage{graphicx}
\usepackage{pgfplots}
\usepackage{tikz} 
\usetikzlibrary{arrows,decorations.pathmorphing,backgrounds,positioning,fit,shapes} 

\usepackage{rotating}

\usepackage[labelformat=empty]{caption} 

\usepackage{multirow}

\usepackage{hanging} 

\usepackage{comment}

\usepackage[compact]{titlesec}
\titleformat{\section}{\normalsize\bfseries}{\thesection}{1em}{}
\titleformat{\subsection}{\normalsize\bfseries}{\thesubsection}{1em}{}
\titlespacing*{\section}{00pt}{5pt}{0pt} 
\titlespacing*{\subsection}{00pt}{0pt}{0pt} 

\newtheorem{propguess}{Property}
\newtheorem{propsubguess}{Property}[propguess]

\makeatletter
\renewcommand\paragraph{%
   \@startsection{paragraph}{4}{0mm}%
      {-\baselineskip}%
      {.5\baselineskip}%
      {\normalfont\normalsize\bfseries}}
\makeatother

\usepackage{color}
\definecolor{Blue}{rgb}{0.0,0.0,1.0}

\usepackage{fancyhdr}
\begin{document}


\begin{center}
\textbf{A DYNAMIC NETWORK APPROACH TO BREAKTHROUGH INNOVATION}
\end{center}

\begin{center}
\begin{table}[htdp]
\begin{center}
\begin{tabular}{cc}
Russell J. Funk & Jason Owen-Smith\\
University of Michigan & University of Michigan \\ 
500 South State Street, \#3001 & 500 South State Street, \#3001\\
Ann Arbor, MI 48109-1382 & Ann Arbor, MI 48109-1382\\
Tel: (815) 272-1518 & Tel: (734) 936-0700\\
\emph{e-mail:} \href{mailto:funk@umich.edu}{funk@umich.edu} &  \emph{e-mail:} \href{mailto:jdos@umich.edu}{jdos@umich.edu}\\
\end{tabular} 
\end{center}
\end{table}
\end{center}

\begin{center}
\textbf{ABSTRACT} 
\end{center}
This paper outlines a framework for the study of innovation that treats discoveries as additions to evolving networks. As inventions enter they expand or limit the reach of the ideas they build on by influencing how successive discoveries use those ideas. The approach is grounded in novel measures of the extent to which an innovation amplifies or disrupts the status quo. Those measures index the effects inventions have on subsequent uses of prior discoveries. In so doing, they characterize a theoretically important but elusive feature of innovation. We validate our approach by showing it: (1) discriminates among innovations of similar impact in analyses of U.S. patents; (2) identifies discoveries that amplify and disrupt technology streams in select case studies; (3) implies disruptive patents decrease the use of their predecessors by 60\% in difference-in-differences estimation; and, (4) yields novel findings in analyses of patenting at 110 U.S. universities.  \\

\begin{center}
\textbf{ACKNOWLEDGEMENTS} \\
\end{center}
\noindent Support for this research was provided by grants from the Rackham Graduate School and the Interdisciplinary Committee on Organizational Studies (ICOS) at the University of Michigan, the National Science Foundation (grant SES-0545634), and an Alfred P. Sloan Foundation Industry Studies Fellowship in biotechnology. We thank Woody Powell, Jeannette Colyvas, Brian Wu, John Chen, Ned Smith, Mark Suchman, members of the Quantitative Methods Program, Corporate Strategy, and the Economic Sociology workshops at the University of Michigan for helpful feedback as well as audiences at the Brown University Commerce, Organizations and Markets Seminar, the 2012 International Conference on Network Science in Evanston, IL, the 2012 Annual Meeting of the Academy of Management, the 2012 Cyberinfrastructure Days Conference held at the University of Michigan, and the Computation Institute at the University of Chicago. All errors are our own.
\normalsize

\pagebreak


\noindent \emph{Economic progress, in capitalist society, means turmoil\ldots New products and new methods compete with the old products and old methods not on equal terms but at a decisive advantage that may mean death to the latter.}--- Schumpeter (1942: 32)\\

\onehalfspacing

\section{INTRODUCTION}
Innovation and its implications are near ubiquitous components of social scientific explanations for stability and change. Yet, existing conceptual and empirical tools are often ill equipped to address the process by which new discoveries ripple through social, economic, and technological systems to reinforce or destabilize the status quo. Despite 70 years of work to develop Schumpeter's (1942) notion of creative destruction, clear and systematic explanations for why, when, and how particular inventions have differential effects on their environments remain elusive. This paper develops a conceptual vocabulary and associated measures to address that important lacuna. 

We argue that a discovery's effects can be read neither from its characteristics at the time of its creation nor from the magnitude of its later use. Both ideas are important, but miss the crucial point that innovations have effects that unfold over time in environments at least partially comprised of other discoveries. The effects inventions have on  subsequent uses of other knowledge are thus essential. 

This notion is aligned with but underutilized in contemporary social scientific work, most of which builds on two core theoretical insights about the dynamics of innovation. First, innovation is a process of recombination (Hargadon \& Sutton, 1997; Fleming, 2001). New technologies, even breakthroughs, are seldom cut from whole cloth; most emerge as previously existing components are arranged into new configurations. Second, technological development is a path dependent process punctuated by periods of discontinuous change (Teece, 1986; Dosi, 1988; Sosa, 2009). Innovations create skills and capabilities that enable the development of more incremental new products and processes. Consequently, important innovations become widely used until they too are rendered obsolete by subsequent breakthroughs, whose own trajectory of use follows a similar pattern.

These perspectives have proven valuable but each is hampered by limitations. The idea of innovation as recombination, for instance, is theoretically ambiguous. Some new configurations are useful but many are worthless, and there exist few means to distinguish between the two \emph{a priori}. Moreover, studies that foreground recombination have difficulty accounting for dynamics because the particular set of components that constitute a discovery are largely fixed at the time of its introduction. Theories that focus on discontinuities in use are better able to account for dynamics, but their emphasis on the magnitude of use---the impact---of a discovery misses the key substantive distinction between new things that are important because they reinforce the status quo and new things that are valuable precisely because they challenge the existing order. 

Closely tied to these theoretical limitations are issues of measurement. Most empirical measures fall into one of two classes according to the properties they use to evaluate innovations. One class seeks to capture \emph{distinctiveness}---the extent to which an innovation departs from existing knowledge. Researchers often measure distinctiveness by, for instance, identifying patents that bridge previously disconnected technological classifications (Trajtenberg et al., 1997; Fleming, 2001; Gruber et al., 2012). 

Like the concept of recombination itself, measures of distinctiveness are limited because they evaluate an innovation at its point of introduction, and neglect the extent and nature of its subsequent use. In so doing, distinctiveness measures miss the possibility that an innovation's true utility unfolds over time in the context of use, organizational strategies, market changes, and the evolution of fields (Bijker, 1995; Rao, 2009). The possibility that a discovery might languish until conditions are right for its recognition and adoption---and the likelihood that such conditions might result from strategic efforts of inventors and organizations---cannot be addressed by measures that emphasize a discovery's features at its conception.

Another class of measures treats discoveries as variable in \emph{impact}---the extent to which they are later used. Technological discontinuities tend to be high impact because they serve as platforms that stimulate future discoveries, but some high impact innovations are important because they advance an ongoing, vibrant technological trajectory. Although the true impact of an innovation is difficult to measure, citations to papers and patents that report new findings and inventions offer a widely used proxy (Hall et al., 2002). But by measuring only magnitude of use, impact measures miss how a technology is used. Consequently, they obscure the key difference between innovations that \emph{amplify} the use of existing ideas and those that \emph{disrupt} them. This ambiguity makes substantive analysis of the social and economic consequences of significant discoveries challenging.

We conceptualize innovations as additions to preexisting networks of complementary or substitutable components. New discoveries alter a network's evolving structure by changing the way subsequent additions connect. These networks, and the fields they represent, change as new entrants diminish or enhance the visibility of incumbents (Powell et al., 2005). Although our focus here is on networks of interdependent technological inventions, the framework we develop could be applied to any system of related physical or conceptual components that grows over time by the addition of nodes and directed links.

Our approach builds on theories that distinguish between innovations that are competency destroying, in that they render existing skills obsolete, and those that are competency enhancing, in that they improve the productivity or value of existing capabilities (Abernathy \& Clark, 1985; Tushman \& Anderson, 1986; Sosa et al., 2012).  We depart from this prior work in several ways, however. First, like much research on innovation, existing studies of competency-enhancing and competency-destroying discoveries tend to emphasize substantial discontinuities that fundamentally break with existing standards. By contrast, we suggest that the extent to which an innovation influences the status quo is a matter of degree, not categorical difference. Second, our approach decouples the amplifying or disrupting effects of new discoveries from the concrete competencies of the organizations that develop and own them. Although the implications of new discoveries are often felt most forcefully in terms of how they influence the value of incumbents' capabilities, their effects ripple through dynamic networks of prior and subsequent discoveries.

To capture this network model of innovation, we develop a relational measure, \emph{disruptiveness}, that quantifies the extent to which an invention amplifies or diminishes subsequent use of the components on which it builds. Combining our index of disruptiveness with an impact weight yields a new operationalization of \emph{radicalness}. These measures are sensitive to both the positive and negative effects of important discoveries.

We begin by reviewing current approaches to technological change in order to identify core properties of breakthroughs and delineate the scope of our contribution. Next, we derive our measures of disruptiveness and radicalness. Subsequent sections validate those measures in three ways. First, we demonstrate that disruptiveness discriminates among inventions of similar impact by applying the measure to U.S. patents issued between 1976 and 2010. Second, we present short case studies of three high-profile inventions that illustrate core features of disruptiveness and radicalness while attesting to the measures' ability to identify recognized breakthroughs. Finally, we cement claims to validity using difference-in-differences estimation, which shows that patents cited by highly disruptive innovations see a 60\% decline in citations after the disruptive patent issues. 

We establish the usefulness of our approach in inferential analyses of the impact, distinctiveness, and radicalness of 55,322 patents assigned to the 110 most research-intensive U.S. universities between 1976 and 2005. Analyzing the radicalness of these patent portfolios yields novel and substantively important results relative to analyses of impact or distinctiveness. We replicate existing findings about the impact of university patent portfolios that demonstrate diminishing returns to commercial experience and positive effects of increases in R\&D sponsorship from both industrial and federal government sources. Models of radicalness, however, manifest starkly different effects, showing that higher levels of industrial support result in the production of \emph{less} radically disruptive patents on campus. Federal R\&D support, by contrast yields \emph{more} radically disruptive discoveries. 

\section{THEORY AND MEASURE---THE EFFECTS OF INNOVATIONS}
\subsection{Properties of Breakthroughs}

Schumpeter distinguished between innovation and invention by emphasizing the impact of discoveries. Innovations are ideas that spark economic transformations by introducing new materials, technologies, or production processes. Inventions are isolated discoveries that have few social or economic effects (Schumpeter, 1942: 132). Although precise terminology can be inconsistent from author to author, there is wide recognition that a core dimension along which innovations vary is in their degree of use (Hall \& Trajtenberg, 2004; Lanjouw \& Schankerman, 2004). These observations suggest a first property of breakthroughs:

\footnotesize
\singlespacing
\begin{propguess}
\begin{minipage}[t]{5.35 in}
A breakthrough discovery is high \underline{\emph{impact}} when it is widely used
\end{minipage}
\label{propImpact}
\end{propguess}
\onehalfspace
\normalsize

Impact is used both conceptually and as family of measures for evaluating innovations. Counts of citations to articles and books have served for decades as proxies for use and scientific quality (Price, 1965). Metrics like impact factor and the \emph{h}-index also quantify use, but for aggregated sets of documents (Hirsch, 2005). Moreover, virtually all studies of diffusion rely on some notion of use or imitation to account for the spread of information, practices, and products (Griliches, 1957; Angst et al., 2010; Greve, 2011). 

Discoveries also vary with respect to how much they differ from their predecessors. Technological developments are seldom completely new but rather consist ``to a substantial extent of a recombination of conceptual and physical materials that were previously in existence'' (Nelson \& Winter, 1982: 130). Important discoveries often involve the unique combination of disparate materials or the rearrangement of previous combinations using novel relationships (Henderson \& Clark, 1990). For example, Hargadon (2002) describes how engineers at Design Continuum combined Reebok's athletic shoe designs with inflatable splints and IV bag technology from the medical device industry to create the Reebok Pump shoe. By contrast, more routine discoveries typically involve slight variations of extant combinations. These observations suggest a second commonly identified property of breakthroughs:

\footnotesize
\singlespacing
\begin{propguess}
\begin{minipage}[t]{5.35 in}
A breakthrough discovery is \underline{\emph{distinctive}} when it recombines existing components in new ways
\end{minipage}
\label{propDistinct}
\end{propguess}
\onehalfspace
\normalsize

Recombination occurs at the time of discovery. Consequently, distinctiveness has been emphasized by researchers who study the innovation process (Burt, 2004; Aral \& Van Alstyne, 2011). For example, Evans (2010) finds that firms' indifference to theory---but generous financial support---leads academic collaborators to produce more papers with more distinctive combinations of biological terms. Fleming and colleagues (2007) find that collaborations between previously disconnected inventors generate distinctive patents that span rarely combined technology classes.

Both impact and distinctiveness fail to capture core dynamics of innovation. Impact differentiates among discoveries based on the extent of their use but says little about how they relate to prior technologies. Distinctiveness discriminates between more unique and more conventional discoveries based on how they relate to prior technologies, but offers little information about the extent of a given technology's future use. The measures are helpful for assessing the value of an innovation (impact) or examining the products of different discovery processes (distinctiveness). Nevertheless, capturing an innovation's implications for economic growth and technological change requires attending to the effects they have on the use of existing technologies. Based on these observations, we propose the following properties.

\setcounter{propguess}{3}

\footnotesize
\singlespacing
\begin{propsubguess}
\begin{minipage}[t]{5.35 in}
A breakthrough discovery is \underline{\emph{amplifying}} when it increases the rate at which future innovations use the components on which it built
\end{minipage}
\label{propAmp}
\end{propsubguess}
\begin{propsubguess}
\begin{minipage}[t]{5.35 in}
A breakthrough discovery is \underline{\emph{disrupting}} when it decreases the rate at which future innovations use the components on which it built
\end{minipage}
\label{propDisrupt}
\end{propsubguess}
\onehalfspace
\normalsize

Properties 3a and 3b relate to the notions of competency-enhancing and competency-destroying discontinuities (Tushman \& Anderson, 1986), but differ in important respects. First, amplification and disruptiveness capture patterns of component use. This generality is attractive because it allows application to a broad range of phenomenon related to innovation. However,  the implications of the waxing or waning use of particular technologies for the value of firm competencies is an empirical question that we expect to vary from context to context. Second, amplification and disruptiveness, unlike competency enhancement or destruction, are continuous properties whose value can change over time. For example, discoveries that are disrupting may be only modestly so; and, their effects may vary in strength. By contrast, the notion of a competency-\emph{destroying} discovery implies a categorical and permanent break with a previous state of affairs.

\subsection{Measuring the Effects of Innovations}
A dynamic network measure of the importance of a new discovery should have four features. First, it should be \emph{structural} in a network sense. Properties 3a and 3b suggest the measure should integrate the degree to which future inventions that build on a discovery also rely on its predecessors. This `second order' form of impact suggests that the importance of an innovation lies more in how it effects the use of other technologies rather than directly in its own use. Second, the measure should be \emph{dynamic}---able to account for variations across the life of a discovery. Our measure of disruptiveness identifies the extent to which a discovery changes the use of its predecessors over time.  While distinctiveness stays fixed, disruptiveness can shift with changes in science and technology, regulation, markets, and other features of the larger environment. Third, the measure  must be \emph{continuous}---able to capture degrees of amplification and disruptiveness. Continuous measures will prove useful for both analyses of large-scale, radical transformations and smaller-scale, incremental shifts that occur at the level of individuals or organizational routines (Feldman \& Pentland, 2003). Finally, the measure should be \emph{valenced}---able to distinguish between disrupting and amplifying innovations. A valenced measure discriminates among innovations with similar distinctiveness and impact, but very different consequences for the status quo. 

We develop and test a measure with these characteristics using U.S. utility patents. Utility patents cover the majority of patented inventions; they include any new (or improved) method, process, machine, manufactured item, or chemical compound. Although other types of patents exist, most patents granted in the U.S. fall in the utility category---over 90\% in 1999 (Hall et al., 2002). To avoid complexities arising from differences in citation practices across categories, we limit our analysis to utility patents. 

Patent data are valuable for our purposes because they have been used extensively, which facilitates comparisons between our results and existing findings. Studies show that a patent's citation impact is correlated with its social and economic value (Hall et al., 2005; Harhoff et al., 1999). Patent data are also attractive methodologically. Before being granted, all patents must pass through an evaluation process. Examiners review prior art citations with an eye toward ensuring comprehensiveness. Applicants have relatively strong incentives to ensure citation accuracy. Listing irrelevant prior art weakens a patent's enforceability, while failing to acknowledge related work can render a patent invalid. In short, patent citations provide a convenient and generally reliable record of the prior technology upon which a patented innovation builds.\footnote{Nevertheless, patent data are known to have limitations. For example, inventors may not seek patent protection for their ideas. Moreover, the process of patent examination has received criticism from observers who argue that the USPTO grants many patents that fail to meet the requirements of novelty and non-obviousness (Jaffe \& Lerner, 2004; Lemley \& Sampat, 2008). We suspect most poor quality patents will score close to zero on our measures, though this remains a matter for empirical testing. For a more complete discussion of the advantages and disadvantages of patent data, see Griliches (1990) and Jaffe et al., (2002).}

Mathematically, we define disruptiveness as follows.\footnote{Should this paper appear in print, we will make a patent-based implementation of these measures and accompanying C++ code publicly available.}  Consider a tripartite graph $G=(V_1, V_2, V_3, E)$, where $V_1$, $V_2$, and $V_3$ represent three classes of nodes and $E$ denotes the edges in $G$. Nodes are only connected if they belong to different classes. Let $V_1$ consist of a focal patent, $f$, $V_2$ consist a set of prior art, $b$, cited by the focal patent, and $V_3$ consist of a set of forward citations, $i$. Classes $f$ and $b$ are fixed at focal patent's issue date, but $i$ can grow as time passes and the patent accrues new citations. Our objective is to model how a new patent, $i$, attaches to the network defined by the ties (citations) between patents of class $b$ and $f$. Attachment takes place in one of three ways: (1) $i$ can cite the focal patent prior art (class $b$), (2) $i$ can cite the focal patent (class $f$), or (3) $i$ can cite the focal patent and its prior art (both class $f$ and $b$). For a focal patent and vector $\mathbf{i} = (i_1, i_2,\ldots,i_{n-1},i_n)$ of forward citations to that patent and/or its prior art at time $t$, we define disruptiveness as
\begin{equation}
D_t = \frac{\sum_{i}{(-2f_{it}b_{it} + f_{it})}}{n_t},
\end{equation}

\noindent where

\begin{equation}
f_{ij} =  \left\{ 
  \begin{array}{l l}
    1 & \quad \textrm{if $i$ cites the focal patent (class $f$)}\\
    0 & \quad \textrm{otherwise,}\\
  \end{array} \right.
\end{equation}

\noindent and 

\begin{equation}
\hspace{45pt} b_{ik} =  \left\{ 
  \begin{array}{l l}
    1 & \quad \textrm{if $i$ cites any focal patent prior art (class $b$)}\\
    0 & \quad \textrm{otherwise.}\\
  \end{array} \right.
\end{equation}
\noindent Here, $n$ is the number of forward cites in $\mathbf{i}$. The measure ranges from -1 to 1, with positive values representing innovations that are more disrupting and negative values highlighting innovations that are more amplifying. Thus, despite the name `disruptiveness,' both disrupting and amplifying innovations are identified. Note that the measure can be readily calculated for both high- and low-impact patents, i.e., patents that receive very many and very few forward citations, respectively. The measure distinguishes impact from disruptiveness using a structural approach that is sensitive to the addition of new citations to the network.\footnote{For the exploratory empirical analyses presented in this paper, we set $t = 2010$.}  It is not defined for patents that do not cite any prior art and that do not receive any subsequent citations by future patents. In network terminology, these patents are isolates, disconnected from other actors in the system. We set the disruptiveness score of these patents to zero, as they have neither impact nor influence over the use of existing innovations.

Finally, the magnitude of a focal innovation's future use can be added to the measure by introducing a weighting parameter to Equation 1 and eliminating the normalization by $n_t$. Given a weight vector $\mathbf{w} = (w_1, w_2,\ldots,w_{n-1},w_n)$,

\begin{equation}
R_t = \sum_{i} \left ( \frac{{-2f_{it}b_{it} + f_{it}}}{w_{it}} \right ), w_{it} > 0 
\end{equation}

\noindent where $w_i$ is the weight given to patent $i$. For simplicity, we set $w_{it}$ equal to 1 so that each $i$ contributes equally.\footnote{A more dynamic (but also complex) approach could, for example, determine the values of $\mathbf{w}$ according to the issue date of $i$ such that older citations have greater or lesser influence over the measure. Further, $\mathbf{w}$ might be weighted according to the distinctiveness of $i$ in order to emphasize the field enhancing or field opening character of a discovery.} We call this measure `radicalness' to indicate the combination of disruptiveness and impact. The measure differs from disruptiveness by distinguishing among innovations according to their overall effect on a network of interlinked technologies. Disruptiveness captures the direction of an innovation's effects, while radicalness mixes both direction and magnitude.\footnote{Our decision to focus on classes of nodes and binary relations them results in the loss of some structural data. We address these issues more fully in the Appendix.}

Radicalness meets the criteria laid out in section 2.1 above. Specifically, it captures impact through a forward citation weight, characterizes disruptiveness on a continuous scale by measuring interrupted citation flows, discriminates between both disrupting (positive disruptiveness) and amplifying (negative disruptiveness) inventions, and its value can vary dynamically as new inventions enter the network.

\section{DISRUPTIVENESS DISCRIMINATES AMONG HIGH IMPACT PATENTS}
\subsection{Data and Methods}
We test our measure's ability to discriminate among patents of similar impact using a sample that consists of the 9,747 U.S. patents granted between 1976 and 2010 that were cited more than 142 times. During this period, 3,983,089 patents were issued. Thus our dataset represents the top 0.25\% of patents by impact. The basic data were obtained from the Patent Network Dataverse (Lai et al., 2011) and supplemented with data taken from the USPTO. Because all of the top 0.25\%  patents are high impact by definition, we calculate the normalized (un-weighted) disruptiveness measure for each. We also report some population-level statistics for all patents granted during the study window to ensure that our conclusions are not driven by the unique properties of this sample.\footnote{While our approach is conceptually straightforward it can be computationally demanding. For example, there are 269,663,060 pathways connecting the prior art of the top 0.25\% patents and their forward citations and more than 100 billion when we consider the full population. Given the scale of the data and complexity of the networks, we made use of a high performance distributed computing environment provided by our home institution to complete these calculations.}

The disruptiveness scores of both the top 0.25\% sample and population are unimodal with modes of 0, means of 0.09, and roughly approximate a normal distribution. A moderately thicker right hand tail in the top 0.25\% sample suggests that more high impact patents disrupt than amplify the use of their predecessors. Far more patents from the full population have disruptiveness scores near 0. The modal patent has little effect on the use of prior technologies. Both distributions lean slightly to the right, which suggests that modestly disrupting innovations may be more common than slightly amplifying ones.\footnote{Prior research indicates that disruptive innovations are rare and therefore is contrary to the control sample distribution, but much existing literature fails to account for the possibility that disruptiveness is a matter of degree. When interpreting the distribution, it is useful to keep in mind that the disruptiveness measure is only modestly correlated with impact ($r = 0.05$, $p < 0.001$) and expressed on a continuous scale. Given that patents are somewhat costly to obtain and that applicants are legally required to demonstrate how their invention is non-obvious, useful, and novel, this slight distributional skew may be expected.} 

Table 1 displays descriptive statistics and correlations for the top 0.25\% patents. The table clearly suggests that our measure functions as intended. The (small) negative correlation between disruptiveness and firm assignee ($r = -0.01$, $p < 0.05$) suggests that firms tend to produce more amplifying innovations, whereas universities ($r = 0.05$, $p < 0.001$) and government laboratories ($r = 0.06$, $p < 0.001$) generate more disruptive breakthroughs.\footnote{Because few patents are assigned to more than one organization (0.59\% in the sample) we only show results for the first assignee.} These correlations are sensible, as companies tend to specialize in application-oriented development while public sector research organizations often emphasize more basic research (Dasgupta \& David, 1994). 

\begin{spacing}{0.5}
\begin{center}
------------------------------------ \\
Insert Table 1 about here \\
------------------------------------ \\ 
\end{center}
\end{spacing}

Disruptiveness is also positively correlated with acknowledgement of a `government interest' ($r = 0.05$, $p < 0.001$).   Government interest patents typically result from federally funded research programs that have been vetted by peer review and are often oriented toward more fundamental discoveries.  Finally, the small positive correlation between team size and disruptiveness ($r = 0.02$, $p < 0.05$) implies that larger teams may produce more disruptive innovations, a potentially noteworthy addition to research on collaborative, team science (Wuchty et al., 2007; Singh \& Fleming, 2010).

\begin{spacing}{0.5}
\begin{center}
------------------------------------ \\
Insert Figure 1 about here \\
------------------------------------ \\ 
\end{center}
\end{spacing}

Figure 1 shows the within year variability of patent disruptiveness for the full population. Consistent with critiques about the exploding volume of patenting activity and perceived lowering of quality thresholds required for obtaining a successful grant, the figure suggests that the average high impact patent of today tends to be notably less disrupting than those of the late 1970s and early 1980s (Jaffe \& Lerner, 2004). At the same time, more outliers score disproportionally high on the continuum of amplifying to disrupting innovation in later years. These observations also manifest in negative correlations between disruptiveness, application ($r = -0.21$, $p < 0.001$), and grant year ($r = -0.21$, $p < 0.001$) among the top 0.25\%. Also of note are the small associations between impact, application ($r = -0.07$, $p < 0.001$), and grant year ($r = -0.06$, $p < 0.001$).\footnote{In unreported  comparisons, we found few substantial differences in the distribution of disruptiveness across broad (NBER) technology categories.  Computers and Communications patents (including software), however, appear to do relatively less to either amplify or disrupt than do other classes.  }

\subsection{Qualitative Assessments of Validity}
Quantitative descriptions of disruptiveness suggest the measure is consistent with established findings about the correlates and features of various types of innovations. But how well is it able to identify and classify particular breakthroughs? Table 2 reports values for the components of disruptiveness and the measure itself for select discoveries that have exerted significant effects on their respective industries. 

\begin{spacing}{0.5}
\begin{center}
------------------------------------ \\
Insert Table 2 about here \\
------------------------------------ \\ 
\end{center}
\end{spacing}

Several of the most amplifying innovations are enhancements in oil and gas drilling technologies. These patents are all assigned to large petroleum companies---the incumbents in an industry with high entry barriers (Orr, 1974). These discoveries do not establish new methods; rather, they refine already widely used technologies. For example, patent 4,573,530, ``In-Situ Gasification of Tar Sands Utilizing a Combustible Gas,'' assigned to Mobile Oil Corporation, describes an improved process for extracting carbon monoxide and hydrogen from tar sands. Such process innovations have become important in recent years because conventional gasification methods are difficult to use in remote regions like Northern Alberta, Canada, where the largest tar sands deposits are located.

The measure also effectively identifies known disruptive innovations. Patents 4,237,224, 4,399,216, and 4,683,202, the Cohen-Boyer patent on recombinant DNA, the Axel patent on the eukaryotic cotransformation, and the Mullis patent on polymerase chain reaction (PCR) appear at the bottom of Table 2.\footnote{By convention, we refer to these patents using the last names of their authors. The inventors' full names are, respectively, Stanley N. Cohen and Herbert W. Boyer, Richard Axel (with Michael H. Wigler and Saul J. Silverstein) and Kary B. Mullis.} These three discoveries from the late 1970s and early 1980s set the stage for the molecular biology revolution in pharmaceutical R\&D by making techniques for targeted in vitro drug discovery possible (Powell \& Owen-Smith, 1998). That technological shift fundamentally altered the structure of the international pharmaceutical industry in the 1980s and 1990s by challenging more traditional, organic chemistry based drug discovery methods (Henderson \& Cockburn, 1996; Powell et al., 1996). These process patents were also exceptionally lucrative for their assignees. The Cohen-Boyer patent generated some \$255 million in licensing royalties for Stanford University over its lifetime (Hughes, 2001). A more aggressive licensing policy at Columbia University led the Axel patent to yield some \$790 million in revenues (Colaianni \& Cook-Deegan, 2009). Finally, some estimates place the total royalties for the Mullis patent at \$2 billion (Fore et al., 2006).

The scanning tunneling microscope (STM, patent 4,334,993) and the atomic force microscope (AFM, patent 4,724,318) together enabled the development of nanotechnology (Darby \& Zucker, 2003; Mody, 2011). Both instruments image and move individual atoms on the surface of a material, which allows electronics, medications, and other products to be designed and built atom-by-atom. While both discoveries are disruptive, the AFM is less so than the STM. This relative ranking accords nicely with the technological history of nanotechnology. Although the STM was introduced five years before the AFM, the AFM offered several major improvements, including 3-dimensional renderings and the ability to image living organisms (Youtie et al., 2008). Although the AFM was a radical improvement over prior generations of microscopes, it was ultimately less disruptive because it built on (and cited) the more transformative STM.

Finally, consider patent 5,016,107, for an ``Electronic Still Camera Utilizing Image Compression and Digital Storage.'' This is an Eastman Kodak patent for an early digital camera that employed novel techniques for image processing, compression, and recording on a removable storage medium (Gavetti et al., 2004; Lucas \& Goh, 2009).\footnote{Note that relative to the perceived disruptiveness of digital photography for silver halide imaging, this patent ranks relatively low on the disruptiveness index with a score of 0.14. One explanation for this is that unlike some of the other technologies discussed above, a single patent did not cover the advent of digital photography (Christensen, 1997; Gavetti et al., 2004). In Appendix A, we present a generalization of the disruptiveness measure that may prove useful for estimating the effects such patent families have on the technologies that precede them.}   

The examples described above cover breakthrough innovations from a wide array of industries and technology classes. They demonstrate that our framework distinguishes between disrupting and amplifying innovations, as well as among degrees within each category.

We now turn to a more detailed consideration of three innovations---glyphosate resistant soybeans, a method of ranking of online search results, and a eukaryotic cotransformation technique---that represent different locations on the amplifying--disrupting spectrum. We chose widely studied breakthroughs to emphasize our measure's utility for the analysis of radically amplifying, moderately disruptive, and radically disruptive inventions. Our focus on innovations drawn from three disparate technology intensive sectors (agriculture, information technology, and biotechnology) also suggest that this approach can inform organizational research across a broad range of fields.

\begin{spacing}{0.5}
\begin{center}
------------------------------------ \\
Insert Figure 2 about here \\
------------------------------------ \\ 
\end{center}
\end{spacing}

Consider Monsanto's patent 6,958,436, entitled ``Soybean Variety SE90346.'' This invention clearly illustrates the core premise of an amplifying discovery. The patent describes a genetically engineered soybean that is resistant to glyphosate, an herbicide patented by Monsanto in the 1970s. Glyphosate is the active ingredient in Monsanto's \emph{Roundup} product line, the number one selling herbicide worldwide. In addition to glyphosate tolerance, the seed integrates other desirable plant traits---like improved yield, immunity to various diseases, and resistance to shattering---using many genetic sequences owned in full or in part by Monsanto. Genetically engineered seeds amplify the value of the earlier patented chemical and biological innovations by broadening potential application areas and excluding competitors (Graff et al., 2003; Pollack, 2009). In this particular case, Monsanto has developed a plant variety specifically engineered to resist damage from its market-leading herbicide (Brennan et al., 2000).

The top panel of Figure 2 plots the network of citations between the Monsanto soybean patent, its prior art, and subsequent inventions that have made use of them. The line graph below the figure tracks the patent's annually updated disruptiveness.  In the network diagram, as in the measure, forward citations are divided into three types. Triangles are patents that cite the focal patent's prior art, but not the focal patent itself. Circles represent patents that cite only the focal discovery and not its prior art. Finally, squares indicate patents that cite both the focal patent and its prior art. 

The figure suggests interesting features of this radically amplifying discovery. Notice that although the soybean patent has received 150 total citations, it has never been cited independently of its prior art. The patent's introduction also virtually eliminated independent citations to its prior art (i.e., triangles). The line graph tracking its disruptiveness starts at zero but declines rapidly to approach -1. This pattern implies that the focal patent and the technologies on which it builds are complementary in a way that would not be expected if this discovery had, for instance, opened a new method of plant engineering.
 
Although not directly revealed in the figure, two additional features of the Monsanto soybean patent are suggestive. First, consider raw citation counts. In the five years before the focal patent was granted, the five pieces of prior art received a total of 61 citations, or 2.44 citations per patent per year on average. In the five years immediately following the soybean patent grant, citations to the prior art increased by more than 600\%, to an average of 15.28 per patent per year.\footnote{Of course, this calculation does not account for possible right truncation due to delays between patent application and grant years, and may also be inflated due to increases in the frequency of patenting and citation over time. Below, we present models that control for these and other factors. The results are broadly similar.}

Second, consider the ownership of the antecedent technologies. By the time of application, Monsanto had acquired the firms that owned all but one of the five pieces of prior art cited by the focal patent.\footnote{Patent 5,084,082 is owned by DuPont.}  This observation fits well with theories of technological innovation that argue incumbent firms strive to enhance the value of their knowledge bases (Sosa, 2011). It also suggests interesting possibilities for using this approach to examine corporate decisions about when to litigate competitor's intellectual property (Allison \& Lemley, 1998; Allison et al., 2004). We would predict, for instance, that the most likely patents to be challenged in established industries are those that amplify the technological position of a significant competitor. A similar logic might be used in explanations of merger and acquisition activities in technologically intensive fields, where complementary patent portfolios often loom large. 

Next, we examine the modestly disruptive patent 6,285,999, entitled ``Method for Node Ranking in a Linked Database.'' This invention, more commonly referred to as PageRank, covers the core algorithm used by Google to weight the importance of web pages for display in its search results. Prior to the introduction of PageRank, search engines used a variety of largely ineffective strategies to rank search results (Brin \& Page, 1998). PageRank, which is owned by Stanford University but licensed exclusively to Google, proposed an entirely new method that drew on insights from social network theory to rank web pages by the number of links they receive from other sites. PageRank's disrupting effects are evident in Yahoo!'s June 2000 agreement to outsource its search function to Google. Prior to the implementation of Google's search technology, Yahoo! was the market leading search engine. After the introduction of this disruptive innovation, Yahoo! could only maintain its position by adopting its competitor's technology.

The middle panel of Figure 2 presents the network of citations involving the PageRank patent, its prior art, and subsequent patents that made use of either. The figure is particularly instructive in comparison to the pattern of citations for the Monsanto soybean patent above it. Unlike the soybean patent, which quickly garnered a substantial number of citations, PageRank appears to have been overlooked in years immediately following its 2001 publication.\footnote{The PageRank patent also differs from the Monsanto case in that its backward citations come from a more diverse base of organizations, as would be anticipated for a more disrupting discovery. The seven pieces of prior art cited by the patent are owned by a total of four different corporations and one university, none of which are Stanford or Google.}  Most citations between 2002 and 2007 cite PageRank's prior art, but not the patent itself. This pattern begins to change between 2007 and 2010, when many more subsequent discoveries cite only the PageRank patent without citing its prior art. It is during this latter period that the PageRank patent began to appear increasingly disruptive to the use of its predecessor technologies.  This pattern of change is also evidenced in the disruptiveness teendline, which begins near zero and rises fairly smoothly to a high of 0.37.

We suspect that the eventual effect of PageRank had as much to do with Stanford's decision about how to license the patent, with the `build fast and monetize later' strategy of growth characteristic of Google's high profile venture capitalists, and with the evolution of the company itself, as it did with technical features of the discovery (Levy, 2011). In other words, this technology's substantial effects may depend intimately on the strategic and tactical decisions of organizations that owned, funded, and developed it. A time-varying version of disruptiveness, such as that illustrated in Figure 2's trend lines, may thus prove useful for theory and analysis of the co-evolution of technologies, industries, and organizations (Murmann, 2003). 

For our final case, we turn to the Axel patent on eukaryotic cotransformation, in the bottom panel of Figure 2. This extremely disruptive discovery is one of the foundational innovations for biotechnology. In contrast to both Monsanto's soybean and Google's PageRank, citations to the prior art cited by Columbia's Axel patent effectively cease within two years following its publication. Out of nearly 340 citations to the Axel patent, only one subsequent invention has cited the discovery together with its prior art and only 14 cited the predecessor technologies on their own.\footnote{One important question is whether or not the Axel patent's high disruptiveness score of 0.95 could stem from the fact that it only cited two pieces of prior art. Put differently, if the patent had cited more predecessor technologies, would the score be lower? Among the top 0.25\% patents, there is a modest negative correlation ($r = -0.20$, $p < 0.001$) between the number of backward citations made and a patent's disruptiveness. However, some association between the amount of prior art cited and disruptiveness should be anticipated. By definition, radically new innovations should have fewer predecessors available on which to build.} Indeed, the trend line below this panel suggests that a radical change in disruptiveness happened three years after issue, and was followed by a smooth rise to a maximum value of 0.95.

The juxtaposition of the cotransformation and soybean patents is interesting when viewed in light of research on technology paradigms and industry life cycles (Dosi, 1988; Klepper, 1997). If the Axel discovery and similar radically disruptive innovations laid the foundations for molecular biology in the late 1970s and early 1980s, then subsequent breakthroughs that follow much later but operate within the same paradigm should be more amplifying (and possibly competency enhancing) in nature, as is the case with the Monsanto patent. In other words, tracing changes in the disruptiveness of innovations belonging to particular `lineages' over time might offer new means to evaluate industry evolution and the process by which knowledge paradigms and technology standards coalesce, expand, and are eventually swept away.

\section{VALIDITY AND USEFULNESS IN ANALYSES OF UNIVERSITY PATENTING}
The preceding descriptive and qualitative analyses suggest the dynamic network framework and associated measures of disruptiveness and radicalness we propose have a high degree of face validity and effectively distinguish among similarly high impact innovations that have disparate effects on their fields. This section examines the population of 55,322 U.S. utility patents issued from 1976-2010 to the 110 most research-intensive American universities. These campuses include every institution that has ever ranked among the top 100 nationally by federal obligations for science and engineering research. The dataset covers a broad set of high and low impact innovations and therefore serves as a useful setting in which to test the value of our approach for substantive analysis at the patent and the organizational levels.

After describing our academic patenting dataset, we use difference-in-differences estimation to examine the effect disruptive patents have on rates of citation to their prior art. Next, we turn to a university-level analysis of the organizational sources of important academic patents. Research on this topic to date has yielded mixed results. We believe the ambiguity in this line of work stems from a reliance on impact measures of innovative importance. Here, we demonstrate that analyzing campus-level sources of more radical patent portfolios offers clear insights into the dynamics of academic patenting that are less apparent in examinations of patent volume, impact, or distinctiveness.

\subsection{Descriptive Statistics and Correlations: Academic Patents}
Table 3 presents descriptive statistics and correlations at the university level for variables that will be used in our inferential analyses.

\begin{spacing}{0.5}
\begin{center}
------------------------------------ \\
Insert Table 3 about here \\
------------------------------------ \\ 
\end{center}
\end{spacing}

The average disruptiveness for academic patent portfolios is 0.10, virtually the same as the average (0.09) reported in Table 1 for both the top 0.25\% sample and the population of utility patents. The standard deviations are very different across the university and high impact samples, however (0.11 and 0.31 respectively). Although most patents are slightly disrupting, when we do not sample on impact, a much larger proportion cluster near zero.

Academic patents also exhibit somewhat different patterns of correlation among the measures of importance. Notably, disruptiveness is more correlated with impact for academic inventions. At 0.16 ($p < 0.001$), however, the association remains relatively slight. Impact is more strongly correlated with university level patent volume ($r = 0.67$, $p < 0.001$). Increasing patent impact may be in part a game of numbers. Disruptiveness exhibits a minimal, non-significant association ($r = -0.03$) with patent volume, suggesting again that our framework addresses features of innovation that are orthogonal to commonly used approaches. Correlations between disruptiveness and measures of university experience with technology transfer, engagement with industry, volume and impact of scientific articles, and federal (National Science Foundation [NSF] as well as National Institute of Health [NIH]) funding are also low, with none rising above 0.13. 

\subsection{Difference-in-Differences Tests at the Patent Level}
We derive difference-in-differences estimates of the effect disruptive discoveries have on the use of their predecessors using a sample of prior art cited by disruptive academic patents and a matched control group of prior art patents cited by inventions with random disruptiveness. The control sample was selected using coarsened exact matching (Iacus et al., 2011; Singh \& Agrawal, 2011). From the 54,322 university patents we selected all those that (1) were one standard deviation above the mean on disruptiveness (among those patents with positive disruptiveness), (2) cited at least one piece of prior art, and (3) fell into one of the six technology categories tracked by NBER (Hall et al., 2002). We then collected the prior art citations for the resulting 2,980 patents. Next, the prior art--focal patent pairs were matched to a set of control prior art--focal patent pairs using (1) the NBER classification of the prior art and focal patent, (2) the grant year of the focal patent, (3) the separation between the grant year of the focal patent and its prior art (matched on deciles of 0-2, 3, 4, 5, 6, 7, 8, 9-10, 11-12, and 13+ years), (4) the total number of citations received by the focal patent's prior art in the three years before and including its issue (matched on deciles of 1, 2, 3, 4, 5, 6-7, 8-10, 11-16, 17-45, and 46+ citations), and (5) the total number of prior art citations made by the focal patent (matched on quintiles of 1, 2, 3, 4, 5, 6-7, 8-10, 11-14, and 15+ citations). A total of 31 focal patents could not be paired with a suitable control, leaving 2,949 patents. Finally some patents were dropped because they (1) only cited prior art granted before 1976 (the year in which our annual citation data begins), (2) were granted after 2006 (and thus did not have sufficient time to influence citations to their prior art) or (3) they did not cite any prior art that fell into one of the NBER technology categories. This step resulted in the loss of 1,112 patents, for a final analysis sample size of 1,837 disruptive patents. These 1,837 patents were then associated with their prior art and matching control patent--prior art pairs; both groups had 2,746 pairs, for a total of 5,492. The data were converted to panel form in order to record annually updated citation counts to the prior art. Each sample had 59,374 focal patent--prior art--years, for an effective panel size of 119,486.

\begin{spacing}{0.5}
\begin{center}
------------------------------------ \\
Insert Table 4 about here \\
------------------------------------ \\ 
\end{center}
\end{spacing}

The result of this procedure is a control sample that closely matches our destructive (treatment) academic patent--prior art pairs. Table 4 reports descriptive statistics for patents in both groups. If disruptiveness validly captures the extent to which an innovation diminishes future uses of technologies that preceded it, then we should see a significant decline in citations to the treatment group of prior art patents relative to the control. This is exactly what we find. 

\begin{spacing}{0.5}
\begin{center}
------------------------------------ \\
Insert Figure 3 about here \\
------------------------------------ \\ 
\end{center}
\end{spacing}

Figure 3 tracks mean citations to our matched samples of treatment and control patents in the five years prior and subsequent to the issuance of a disruptive patent. The $x$-axis is a timeline standardized around the year (0) in which focal treatment or control patents were granted. One year before `treatment' ($x$-axis = -1) the prior art patents in both samples were cited at an almost identical rate of 0.78 (treated) and 0.87 (control) citations. Three years later---two years after treatment ($x$-axis = 2)---a gap opens. Citations to control patents (the dashed line) have risen to a peak of 1.24 citations while citations to the prior art of disruptive innovations have declined to 0.88, on average. Thus, disruptiveness identifies technologies that substantially interrupt subsequent use of discoveries on which they built.

The inset table reports differences across groups of patents and time periods. The first column summarizes rates of citation to treatment and control patents in the five-year period before a disruptive patent issues. These well-matched samples are cited at a remarkably similar rate, with treated patents receiving only 0.04 fewer citations on average. After a disruptive patent issues, however, treated patents are cited on average 0.33 fewer times than control. Citations to both samples grew in the later period, but citations to treated patents grew at a rate that was 0.29 lower ($p < 0.001$), a decline of some 60.1\% relative to the control.\footnote{For details on significance tests for the difference-in-differences estimation, see Figure 3.}

Thus far, our evaluation of disruptiveness and radicalness has addressed the face validity and discriminatory power of our framework relative to more common approaches that rely on indicators of impact and distinctiveness. We have demonstrated that (1) disruptiveness is only modestly correlated with impact, (2) the discoveries our approach identifies as disrupting or amplifying are substantively sensible and fit well with published evaluations of high profile innovations, and (3) the prior art cited by disruptive patents exhibits a substantial decline in citations relative to a closely matched control group. In other words, the measures appear to validly operationalize the theoretical concepts they target and accurately characterize the effects disparate types of discoveries have on the use of their predecessors. But, is being able to quantify these theoretically interesting differences likely to lead to new substantive findings? Put simply, are these new measures useful research tools?  We address this question using disruptiveness and radicalness as dependent variables in models that explore several competing arguments about the factors that contribute to universities' ability to generate important patents.  

\subsection{Research on Academic Patenting}
In 1980, Congress passed the Bayh-Dole Act; a law that allows organizations that perform federally funded research to file for patents and issue licenses on intellectual property (IP) they develop. The act accelerated a trend toward research commercialization on campus. Over the last thirty years, academic patenting has increased dramatically. Some inventions have been highly lucrative for the institutions that own them by generating new products, companies, and even industries.

Proponents of Bayh-Dole call it `prescient' (Cole, 1993) and herald the act's importance in turning university science into an engine of economic development (Powell et al., 2007). In 2002, \emph{The Economist} (2002: 3) called the act ``possibly the most inspired piece of legislation enacted in America over the past half century'' and went on to attribute it an important role in ``reversing America's precipitous slide into industrial irrelevance.'' By the same token, critics of Bayh-Dole and commercialization bemoan the `selling' of the university (Greenberg, 2007), link the act to the `corporate corruption' of academe (Washburn, 2005), and argue that proprietary research diminishes universities' scientific and public mission (Krimsky, 2004).

Hyperbole aside, both the economic benefits and the dangers of academic commercialization are often attributed to the type of inventions that universities produce. For those with rosier views, it is precisely the university's capability to generate unexpected, market-creating inventions outside the channels of corporate R\&D that make academic innovations valuable. In the terms we use here, the economic benefits of Bayh-Dole rely in part on academic scientist's propensity to generate radically disrupting inventions. Although critics approach Bayh-Dole from a range of philosophical starting points, one thread running through most negative appraisals is concern that attention to the commercial value of academic ideas leads to capture by corporate interests, and with it to university science closely wedded to industry priorities. Put differently, one significant concern about Bayh-Dole suggests that as research commercialization becomes more widespread, both published and patented science will tend to amplify the status quo. 

This tension has led many academic analysts to assess the costs and benefits of university research commercialization (Trajtenberg et al., 1997; Henderson et al., 1998; Mowery et al., 2002). Two key themes in this line of work emphasize (1) the value of academic innovations, and (2) the relationship between increases in the use of proprietary science and the vitality of more fundamental research and training. 

Scholars pursuing the former question express concern over the relationship between the sources and types of R\&D funding that support academic research and the volume or impact of patented science on campus.  Although the results of some studies conflict, most find evidence that increases in the quantity of academic patenting do not decrease its impact at the campus level (Mowery et al., 2002). These studies also suggest that connections with industry help academic institutions learn to patent up to a point. Owen-Smith and Powell (2003) find a curvilinear (inverted-U shaped) relationship between ties to biotechnology firms and the impact of a university's patent portfolio. They attribute diminishing returns to the likelihood of corporate capture of technology transfer priorities.

The relationship between academic (papers) and proprietary (patents) research outputs has also been much examined.  Owen-Smith (2003) demonstrates that patent flows are deeply intertwined with traditionally academic inputs and outputs such as federal grants and publications. Sine et al. (2003) find that academic visibility in the form of article citations has increases the likelihood patents will be licensed by industry. For a set of life science patents issued to 89 university campuses, Owen-Smith and Powell (2003) find a positive relationship between the citation impact of life science papers and the overall impact of a university's patent portfolio.

\textbf{Technology Transfer Experience.} Prior work reports mixed results about the effects of experience with patenting. More experience in the form of increasing patenting may be associated with declines in impact, as universities looking to develop strong patent portfolios pursue more incremental innovations (Henderson et al., 1998).  An alternative proxy for experience, the age of a university's technology transfer office, suggests different relationships. Most technology transfer units function at an economic loss, yet many are under intense pressure to generate revenue (Kenney \& Patton, 2009). The result is often that the youngest and most tenuous offices are more likely to pursue quick and relatively sure innovations that are modestly amplifying and thus easily marketed to potential industrial licensors, rather than more difficult to evaluate disruptive inventions that might be harder to sell and slower to realize economic returns. In sum, the literature leads us to expect that increasing experience with technology transfer will also increase the impact of university patent portfolios, but that same experience might plausibly lead campuses to produce either more amplifying or more disruptive patents.

\textbf{Research Support.} Owen-Smith \& Powell's (2003) observation that greater corporate engagement leads to the capture of university technology transfer efforts suggests that science supported by industrial partners is likely to generate innovations that increase the use of existing technologies. We thus propose that industry support of university R\&D will lead universities to pursue more amplifying discoveries. Likewise, levels of support from the NSF and NIH, the nation's premier funders of basic science, should yield research that is less connected to existing industrial needs. Thus, more federal grants and more patents derived from them should be associated with the generation of more disruptive innovations.

Broadly speaking, if the decentralized peer-reviewed process by which federal science agencies apportion R\&D support are more likely to fund research focused on academic concerns divorced from the needs of industry, then campuses that perform more federally funded research and those whose patents emerge from federally funded projects should produce more radically disruptive patents. Likewise, because corporate R\&D funding is presumably linked to sponsors' proximate, market-driven goals, we expect campuses that pursue more industrially funded R\&D to produce patents that amplify the use of existing technologies. While any source of external support seems likely to yield higher impact patents, we hypothesize that industrial connections will be associated with amplifying innovations, while public sector funding will lead to more disruptive discoveries.

\textbf{Scientific Capacity.} Although existing measures of publication impact suffer from many of the same limitations as the patent citation measures we critique above, we use them to provide some sense of the relationship between high-impact public and important proprietary science. We follow two lines of reasoning in our analysis. Noting first with Sine and colleagues (2003) that articles confer a `halo effect' on associated academic patents while advertising their value to potential licensees, we expect both the volume and impact of academic science to be associated with higher-impact patent portfolios. To the extent that highly-cited papers generate scientific attention because they report novel, basic-science discoveries that are far removed from the current concerns of industry, we would expect higher-impact publications to be associated with more radically disruptive innovations.

\subsection{Model Estimation and Results}
We first model patent volume as a simple yearly count of issued patents by application date. To capture time-invariant heterogeneity among universities and to account for overdispersion, we use a conditional fixed effects negative binomial specification (Hausman et al., 1984). We next consider the impact of academic patent portfolios using a count of forward citations. Because most patents require several years to accumulate the bulk of their citations, we count citations from the year of application---with the earliest being applied for in 1981---to 2010. This means that patents applied for at the end of our panel in 2005 have a full five years to accumulate citations, which should attenuate right censoring bias resulting from citation lags (Hall et al., 2002). Here too we mobilize a fixed effects negative binomial specification.

Our next dependent variables operationalize the idea of distinctiveness using Trajtenberg and colleagues (1997) index of originality and Fleming's (2001) measure of technology area combinations. Both constructs emphasize the extent to which new innovations reach across existing technology categories at the time of their application. The former is a Herfindahl measure of the primary classes of the patents a focal patent cites; the latter is an indicator variable that takes on a value of 1 if a focal patent bridges a previously uncombined set of secondary technology classes and 0 otherwise. Finally, we estimate models of disruptiveness and radicalness, which are both calculated using citations received between the application year and 2010 so that the measures parallel our impact variable. We aggregate these measures to the portfolio level by averaging across all patents applied for by each university at time $t + 1$. Because all of these last four variables are continuous, we adopt an OLS specification that includes year and university fixed effects.

Our independent variables correspond to the sets of explanations outlined above. Table 5 displays the coefficient estimates for each dependent variable.\footnote{Tables with nested model specifications for each variable are available upon request.}  We discuss our findings from Table 5 row-by-row to emphasize the disparate effects of the same independent variables on different outcome measures. 

\begin{spacing}{0.5}
\begin{center}
------------------------------------ \\
Insert Table 5 about here \\
------------------------------------ \\ 
\end{center}
\end{spacing}

\textbf{Technology Transfer Experience.} The first four rows of Table 5 report the effects of variables that index a campus's experience with technology transfer. Patent stock reflects a university's aggregate experience. Increased patenting experience is associated with greater yearly patent flows (Model 1) but not with increased impact (Model 2), distinctiveness (Models 3 and 4) or disruptiveness (Model 5). Model 6 documents a strong negative relationship between patenting experience and radicalness.  As universities succeed in their pursuit of more IP, their patents tend to amplify existing technological arrangements.

The age of a university's technology transfer office and its quadratic term also demonstrate that the relationship between experience and patent quality is more nuanced than suggested by previous studies. Age has an inverted U-shaped relationship with both patent impact (Model 2) and radicalness (Model 6); however, there is a negative and linear relationship between this measure of experience and disruptiveness (Model 5). The relative size of first and second order coefficients in Models 2 and 6 suggest that there are diminishing returns to experience rather than a reversal of effects. 

The strong negative coefficient reported in Model 5 implies that experienced offices pursue patents on more amplifying innovations. This may stem from several mechanisms. First, established technology transfer offices might pursue repeated licenses with the same corporate partners, or use a small number of well-known partners to help vet disclosures (Owen-Smith, 2005). Either activity could closely link a university's patent portfolio to partners' proprietary concerns, resulting in more amplifying IP. Alternatively, established offices may strategically pursue patents on suites of related technologies or because of the important role that long term relationships with a small number of prolific inventors tend to play in academic technology transfer (Colyvas, 2007). Either dynamic would result in more amplifying patents, as universities work to protect families of innovations built from interdependent IP. 

\textbf{Scientific Capacity.} We use the log of scientific articles published in the preceding year to index the volume of academic research on campus. In addition, we include a year and subject area standardized measure of the citation impact factor of those articles as a proxy of the visibility and quality of a campus's published research. Both measures are drawn from the Institute for Scientific Information's (ISI) University Indicators Database. 

Model 2 shows positive and significant effects of impact factor and quantity of scientific articles on overall patent impact. Thus, consistent with prior research, the quality and quantity of an institution's science signals value to industrial interests. Models 5 and 6 show significant, positive effects of impact factor on patent disruptiveness and radicalness, respectively, which suggests that institutions with higher quality science tend to produce more disruptive patents. We do not find a significant association between the quantity of scientific articles and radicalness. Taken together, these findings provide noteworthy support for the idea that highly visible academic research yields both higher impact and more disruptive patents. 

\textbf{Research Support.} Industry-sponsored R\&D measures the dollar value, in millions, of grants and contracts from corporations to researchers on campus. This variable comes from NSF surveys of campus level R\&D efforts available from the online WebCaspar database.\footnote{\url{https://webcaspar.nsf.gov/}} Our second measure, industry contractual ties, emphasizes formal relationships with firms in technology intensive industries. Information on financial, R\&D and licensing connections were content coded from Securities and Exchange Commission filings made by a sample of 634 publicly traded firms in high-technology sectors (c.f. Buhr \& Owen-Smith 2010). The results reported in Model 2 (impact) and Model 6 (radicalness) are particularly interesting. Higher levels of industrial R\&D support lead to increases in the impact of university patent portfolios while decreasing their radicalness, a pair of findings consonant with the idea that corporate partners support high quality research that is related to an existing technology base. Industrial R\&D support may thus have the effect of more tightly linking the expertise and concerns of academic researchers with corporate funders, resulting in patents that strengthen the position of incumbent firms. Having more contractual ties to corporate partners likewise diminishes the radicalness of university innovations (Model 6), but that finding may result from the fact that universities that are more closely connected to industry also generate lower-impact patent portfolios. Neither measure of industry connection has a significant effect on disruptiveness.

Finally, we consider how federal funding for science and engineering helps shape the importance of academic patents. We track the level of support (in millions of dollars) that flows to campuses from the two premier federal science agencies, the NSF and the NIH. Additionally, we measure the number of patents in a university's portfolio that include an acknowledgement of `government interest.' Patents that emerge directly from federally funded research projects must allow the state to claim certain ownership rights as a result of that funding. Thus, universities with a greater number of government interest patents have patent portfolios that are more closely linked to federally funded R\&D activities. Federal grant measures were extracted from the NSF's WebCaspar database, while information on government interest was taken from full text patent documents. Neither NIH nor NSF funding by itself is related to the impact of university patent portfolios, whereas increases in both lead campuses to patent more radically disruptive innovations. Government interest patents, by contrast, are positively associated with both impact and radicalness. 

The different effects of industrial and federal R\&D support in these models are stark. More industry connections are associated with the production of more amplifying patents.  Federal support yields more disruptive IP.  Both are associated with higher impact. Thus, impact is an ambiguous construct that does not take into account the key substantive differences between inventions whose value derives from bolstering the status quo and those that are important because they challenge it.

Our analyses hold lessons for both defenders and detractors of Bayh-Dole. For the former, the story is clear: the sometimes-spectacular wave of university inventions (including those we highlight in Table 2) depends on the foundation of academically important, publicly funded science. Thus, to preserve their economic contributions, the academic character of university science should be encouraged. For the latter, it is apparent that tighter industrial connections yield patents that reinforce existing technical arrangements; a finding that should deepen concerns about the long-term effects of commercialization on universities, their research, and ultimately their economic contributions. 

\section{DISCUSSION}
The dynamic network approach and accompanying measures of innovation disruptiveness and radicalness we propose capture theoretically important, but thus far underutilized, information about the evolution of networks of related inventions. Those structural, dynamic, continuous, and valenced measures discriminate among the highest impact patents, identify some of the most high profile disruptive and amplifying inventions of the last three decades, and suggest many new avenues of research in a variety of fields. When applied to substantive questions about the costs and benefits of university research commercialization, the package of theoretical and empirical tools presented here yields starkly different results on some dimensions, while replicating current knowledge on others. Both sorts of findings have important implications for ongoing academic and policy debates. 

We began by noting that although canonical theories of innovation and technological change emphasize the effects that new discoveries exert on the value and utility of their predecessors, available measures do little to address these dynamics. These measurement shortcomings can lead to empirical findings that are ambiguous and contradictory, limiting social scientists' ability to develop and refine theories of innovation. 

Disruptiveness does an effective job of distinguishing among important innovations. Among the 0.25\% most cited patents, the measure is roughly normally distributed and nearly uncorrelated with relative differences in impact ($r = 0.05$, $p < 0.001$). Our measure also holds up well to descriptive tests of face validity in both aggregate terms and as pertains to individual, high profile discoveries. In short, disruptiveness appears to be capable of distinguishing the right kinds of innovations from each other on both qualitative and quantitative dimensions. Disruptive patents cause a dramatic decline in citations to their prior art. The measure is orthogonal to impact and captures precisely the empirical effect that is its target. Finally, radicalness yields clearer findings about the institutional and organizational sources of academic patents than do more standard measures. These findings suggest novel implications for the study of innovation in this important realm. 

Below, we briefly address implications of our perspective for a variety of substantive areas where it is likely to be particularly useful. 

\textbf{Implications for Science Policy.} In recent years, efforts to develop evidence-based science policy measures have triggered extensive work in what has come to be called the science of science policy (Lane, 2010; Fealing et al., 2011). Among other concerns in this burgeoning field is the question of how to support, identify, and evaluate transformative research. We submit that variants of the measures we propose might aid in efforts to rigorously document the transformative effects of different collaborative or funding models by, for instance, associating differing levels of interdisciplinarity or scientific team structure with the production of more or less radical findings that enhance or replace existing knowledge bases.

\textbf{Implications for Research on Social Networks and Innovation.} The dynamic network approach we propose here may also help generate new insights in research on social networks and innovation (Hansen, 1999; Burt, 2004; Obstfeld, 2005). A persistent challenge in this area has been to distinguish between the effects of more open, unconstrained social networks and more closed or cohesive structures. Current findings suggest that the former are more effective for generating new ideas while the latter are more suited to the oftentimes-difficult work of development. The distinction between discovery and implementation, though, may mask a more fundamental difference between network positions that facilitate novel but disruptive innovations and those that support more amplifying discoveries. 

Perhaps more significantly, the commonplace distinction between cohesive networks and more open structures may be insufficient for explaining the different effects of similarly good ideas. Search through cohesive ties, for instance, might yield amplifying discoveries, while brokerage in open networks may be associated with innovations that have the potential to disrupt the status quo of the context from which they spring. The dynamic network framework we propose may also be useful for identifying when brokers are bridging groups engaged in different, though jointly amplifying, lines of work, and when the technological streams they seek to develop are at odds or simply orthogonal to one another. 

\textbf{Implications for Economic Geography.} This approach could also be of use in work on the causes and consequences of regional industrial agglomeration in high technology industries. Competition is stiffer within clusters than outside them (Stuart \& Sorenson, 2003). Recent work has documented the different roles that physical proximity to universities and other firms and geographically local versus distant ties play in the innovative performance of biotechnology companies (Owen-Smith \& Powell, 2004; Whittington et al., 2009). Both lines of work suggest that location in vibrant regions forces firms to compete more vigorously, but research in this area has not attempted to distinguish among the types of discoveries made by geographically clustered and geographically isolated organizations. To the extent that such clusters constitute centers of gravity for many high technology industries, their implications for the amplifying or disruptive characteristics of discoveries would be of interest to scholars, managers, and policymakers.

\textbf{Implications for Studies of Technological Change. } Measures of radicalness and disruptiveness will also prove useful in efforts to understand the maturation and disappearance of technological standards (Nelson \& Winter, 1982; Dosi, 1988; Klepper, 1997). Recall our brief discussion of two related technologies---the Scanning Electronic Microscope (SEM) and Atomic Force Microscope (AFM)---identified in Table 2. In our view, the SEM is a greater breakthrough innovation because it was more disruptive. The AFM, in contrast, was only moderately disruptive because it built on and expanded the technological base created by the SEM. Yet, the less disruptive AFM has proven to be more important to the development of nanotechnology. This dynamic suggests interesting possibilities for studying the evolution of new technology areas such as nanoscale materials (Darby \& Zucker, 2003) or tissue engineering (Murray, 2002). In both these arenas, we might expect early innovations to disrupt incumbent's capabilities, but as the technologies (and related organizations, networks, and markets) mature we might also find later discoveries shifting toward the amplifying end of the spectrum. Studies of punctuated technological and scientific change may benefit from such analyses if disruptive innovations are followed by successively more amplifying discoveries that build on and extend their reach. More mature fields dominated by less destructive, more incremental innovation may also be ripe for radical discoveries that will spark another cycle of technological development. 

\textbf{Implications for Research on Corporate Strategy.}  The ability to distinguish between discoveries that amplify and those that erode the use of existing technologies may also be valuable for examining a range of questions in corporate strategy research. For instance, consider debates over the effect of market power on incentives to innovate. Much like research on university patenting, this literature has produced conflicting theoretical arguments and empirical findings (Gilbert, 2006; Ahuja et al., 2008). Although some scholars contend that incumbents should pursue innovation in order to maintain their positions (Schumpeter, 1942; Christensen, 1997), others argue the opposite and claim that incentives are depressed due to lack of competition (Arrow, 1962). Ahuja and colleagues (2008: 8) note that empirical efforts to resolve this debate have been challenging, ``since the main effects are countervailing.'' To the extent that firms may differ with respect to the amplifying or disruptive aims of their R\&D efforts, the framework we propose here could help to disentangle some of these countervailing effects.

\textbf{Implications Beyond Innovation Research. } Finally, consider some uses of a dynamic network approach to innovation far afield from the world of technological change. Our disruptiveness and radicalness measures might profitably be adapted to studies of the evolving link structure of the World Wide Web, where new pages will cite and be cited by others. Similarly the evolving structure of citations among judicial decisions that comprise an important aspect of U.S. law might profitably be examined in these terms. Rather than measuring importance ultimately in terms of competency destruction or enhancement, these measures might offer new insight into more political and cultural questions involving the dynamics of polarization in the blogosphere (Adamic \& Glance, 2005), the evolution of management fads and fashions (Strang \& Macy, 2001), the visibility of music artists and genres (Rossman et al., 2008), the commercial and critical impact of films, and the larger legal implications of new court decisions. Evolving network data that offer insights into relationships of deference among entrants and incumbents in a variety of substantive domains are becoming more readily available and are important to research questions in many disciplines. In this context, the dynamic network approach we propose has the potential to serve countless useful purposes.

\section{CONCLUSION}
We began this article with a foundational quote from Schumpeter. That insight served as a stepping stone for nearly two generations of research that took the notion of innovation-based creative destruction as a motor for economic, technological, organizational, and social transformation. While Schumpeter and many others have noted that what makes a given invention important is the extent to which it is used, the question of how, precisely, such uses strengthen or challenge the status quo has remained elusive. This article outlines a dynamic network approach to the study of breakthrough innovations that foregrounds the effect new discoveries have on the uses to which their successors put the technologies upon which they depended. In so doing, we more fully operationalize a key but as yet under-examined component of most theories of innovation: the fundamental distinction between innovations that are valuable because they are disruptive of current standards and those whose value derives from the amplification of the trajectories from which they spring. While not without limitations, we believe this approach and the measures we present and validate have substantial, exciting possibilities to expand research in multiple fields. 

\singlespacing
\section{APPENDIX}
For the purposes of presentation, we simplified the disruptiveness index in several ways.  Specifically, the variant of the measure used in the text is limited in that it (a) assumes that there is only one one node in the focal class (i.e., one focal patent), and (b) for each node in a given class, collapses what may be multiple cross class ties into a single 0, 1 indicator. In this appendix, we present a more general version of the measure that removes these simplifications.  

Consider a tripartite graph $G=(V_1,V_2,V_3,E)$.  Let $V_1$, $V_2$, and $V_3$ represent three classes of nodes and $E$ denotes the edges in $G$.  Nodes can only be connected if they are members of different classes.  For illustrative purposes, let $V_1$ represent a set of $q$ pieces of prior art, $V_2$ represent a set of $m$ focal patents, and $V_3$ represent a set of $n$ forward citations. To capture the pattern of citation between classes of vertices, we define two incidence matrices, $\mathbf{F}$ and $\mathbf{B}$, which record links between the forward citations ($V_3$) and the focal class ($V_2$) and the forward citations ($V_3$) and the prior art ($V_1$), respectively, such that   

\begin{equation}
f_{ij} =  \left\{ 
  \begin{array}{l l}
    1 & \quad \textrm{if vertex $n$ of class $V_3$ cites vertex $m$ of class $V_2$}\\
    0 & \quad \textrm{otherwise,}\\
  \end{array} \right.
\end{equation}

\noindent and 

\begin{equation}
b_{ik} =  \left\{ 
  \begin{array}{l l}
    1 & \quad \textrm{if vertex $n$ of class $V_3$ cites vertex $q$ of class $V_q$}\\
    0 & \quad \textrm{otherwise.}\\
  \end{array} \right.
\end{equation}

The incidence matrices can be thought of as analogues to adjacency matrices for multipartite graphs.  Here, the rows of both $\mathbf{F}$ and $\mathbf{B}$ correspond to the elements of $V_3$ (forward citations).  The columns of $\mathbf{F}$ are the elements of $V_2$ (focal patents), while the columns of $\mathbf{B}$ are the elements of $V_1$ (prior art).  Using the information stored in the incidence matrices, we can now define a more complete version of the measure as

\begin{equation}
D^\prime = \sum\limits_{i=1}^n\left[{-2\left(\frac{\sum\limits_{j=1}^m f_{ij}}{m}\right)\left(\frac{\sum\limits_{k=1}^q b_{ik}}{q}\right) + \left(\frac{\sum\limits_{j=1}^m f_{ij}}{m}\right)}\right]\Bigg /{n},
\end{equation}

Dividing each sum by the total number of focal patents ($m$) or pieces of prior art ($q$) available for citing effectively normalizes the measure so that it ranges from -1 to 1. A corresponding version of radicalness can be obtained by eliminating the normalization and introducing a vector of weights $w=(w_1, w_2,\ldots,w_{n-1},w_{n})$ such that

\begin{equation}
R^\prime = \sum\limits_{i=1}^n\left[{\left(-2\sum\limits_{j=1}^m f_{ij} \sum\limits_{k=1}^q b_{ik}  +  \sum\limits_{j=1}^m f_{ij}\right)\Bigg /{w_i}}\right], w_i > 0.
\end{equation}

The ability to expand the size of class $V_2$ (focal patents) may be useful in a variety of extensions, particularly for research on patent pools or corporate intellectual property portfolios.  This variant of the measure is also attractive in that it retains some structural data that is lost through the collapsing of multiple ties into a single indicator as described above. However, it is important to note that this approach gains nuance at the expense of added complexity.  To continue with the example of patents, in order for a set of patents ($V_2$) to be maximally amplifying on this metric, all forward citations would need to cite all members of focal class and all prior art cited by that class.  Similarly, maximal disruptiveness would only be achieved if all forward citations cited all members of the focal class and no prior art.  We may, as a result, expect the values of the measure to cluster more closely around 0 as the size of $V_2$ grows larger.  Additional complications arise from the fact that, at least in the patent context, members of the focal class may cite on another.  If course, this is only a problem if $V_2$ has more than one element.\\

\section{REFERENCES}
\begin{hangparas}{.20in}{1}

Abernathy WJ, Clark KB. 1985. Innovation: Mapping the winds of creative destruction. \emph{Res. Policy} 14:3-22.

Adamic, LA, Glance N. 2005. The political blogosphere and the 2004 US Election: Divided they blog. In \emph{Proc. 3rd Internat. Workshop on Link Discovery}. Chicago, Illinois.

Ahuja G, Lampert CM, Tandon V. 2008. Moving beyond Schumpeter: Management research on the determinants of technological innovation. \emph{Acad. Management Ann.} 2:1-98.

Allison JR, Lemley MA. 1998. Empirical evidence on the validity of litigated patents. \emph{AIPLA Quart. J.} 26:185-275.

Allison JR, Lemley MA, Moore KA, Trunkey RD. 2004. Valuable patents. \emph{Georgetown Law J.} 92:435-479.

Angst CM, Agarwal R, Sambamurthy V, Kelley K. 2010. Social Contagion and Information Technology Diffusion: The Adoption of Electronic Medical Records in U.S. Hospitals. \emph{Management Sci.} 56:1219-1241.

Aral S, Van Alstyne M. 2011. The diversity-bandwidth trade-off. \emph{Amer. J. Sociol.} 117:90-171.

Arrow KJ. 1962. Economic welfare and the allocation of resources for invention. In \emph{The Rate and Direction of Inventive Activity}, edited by HM Groves. Cambridge, MA: National Bureau of Economic Research.

Bertrand M, Duflo E, Mullainathan S. 2004. How much should we trust differences-in-differences estimates? \emph{Quart. J. Econom.} 119:249-275.

Bijker WE. 1995. \emph{Of Bicycles, Bakelites, and Bulbs}. Cambridge, MA: MIT Press.

Brennan, MF, Pray CE, Courtmanche A. 2000. Impact of industry concentration on innovation in the US Plant biotech industry. In \emph{Transitions in Agbiotech}, edited by WH Lesser. Storrs, CT: University of Connecticut.

Brin S, Page L. 1998. The anatomy of a large-scale hypertextual web search engine. \emph{Comput. Networks ISDN Systems} 30:107-117.

Buhr H, Owen-Smith J. 2010. Networks as institutional support: Law firm and venture capitalist relations and regional diversity in high-technology IPOs. \emph{Res. Sociol. Work} 21:95-126.

Burt RS. 2004. Structural holes and good ideas. \emph{Amer. J. Sociol.} 110:349-399.

Christensen CM. 1997. \emph{The Innovator's Dilemma}. Boston: Harvard Business School Press.

Colaianni A, Cook-Deegan R. 2009. Columbia University's Axel patents: Technology transfer and implications for the Bayh-Dole act. \emph{Milbank Quart.} 87:683-715.

Cole JR. 1993. Balancing acts: Dilemmas of choice facing research universities. \emph{Daedalus} 122:1-36.

Colyvas JA. 2007. From divergent meanings to common practices: The early institutionalization of technology transfer in the life sciences at stanford university. \emph{Res. Policy} 36:456-476.

Darby MR, Zucker LG. 2003. Grilichesian breakthroughs: Inventions of methods of inventing and firm entry in nanotechnology. Cambridge, MA: NBER Working Paper 9825.

Dasgupta P, David PA. 1994. Toward a new economics of science. \emph{Res. Policy} 23:487-521.

Dosi G. 1988. Sources, procedures, and microeconomic effects of innovation. \emph{J. Econom. Literature} 26:1120-1171.

Evans JA. 2010. Industry induces academic science to know less about more. \emph{Amer. J. Sociol.} 116:389-452.

Fealing KH, Lane JI, Marburger III JH, Shipp SS. 2011. \emph{The Science of Science Policy}. Stanford, CA: Stanford Business Books.

Feldman MS, Pentland BT. 2003. Reconceptualizing organizational routines as a source of flexibility and change. \emph{Admin. Sci. Quart.} 48:94-118.

Fleming L. 2001. Recombinant uncertainty in technological search. \emph{Management Sci.} 47:117-132.

Fleming L, Mingo S, Chen D. 2007. Collaborative brokerage, generative creativity, and creative success. \emph{Admin. Sci. Quart.} 52:443-475.

Fore J, Wiechers IR, Cook-Deegan R. 2006. The effects of business practices, licensing, and intellectual property on development and dissemination of the polymerase chain reaction: Case study. \emph{J. Biomedical Discovery Collaboration} 1.

Gavetti G, Henderson R, Giorgi G. 2004. Kodak (a): Harvard Business School Case.

Gilbert R. 2006. Looking for Mr. Schumpeter: Where are we in the competition-innovation debate? \emph{Innovation Policy Econom.} 6:159-215.

Graff GD, Gordon CR, Small AA. 2003. Agricultural biotechnology's complementary intellectual assets. \emph{Rev. Econom. Statist.} 85:349-363.

Greenberg DS. 2007. \emph{Science for Sale}. Chicago: University of Chicago Press.

Greve HR. 2011. Fast and expensive: The diffusion of a disappointing innovation. \emph{Strategic Management J.} 32:949-968.

Griliches Z. 1957. Hybrid corn: An exploration in the economics of technological change. \emph{Econometrica} 25:501-522.

Gruber M, Harhoff D, Hoisl K. 2012. Knowledge Recombination across Technological Boundaries: Scientists Vs. Engineers. \emph{Management Sci.}, enline July 18, 2012.

Hall BH, Jaffe AB, Trajtenberg M. 2005. Market value and patent citations. \emph{RAND J. Econom.} 36:16-38.

Hall BH, Jaffe AB, Trajtenberg M. 2002. The NBER patent-citations data file: Lessons, insights, and methodological tools. In \emph{Patents, Citations \& Innovations}, edited by AB Jaffe and M Trajtenberg. Cambridge: MIT Press.

Hall BH, Trajtenberg M. 2004. Uncovering GPTs with patent data. Cambridge, MA: NBER Working Paper 10901.

Hansen MT. 1999. The search-transfer problem: The role of weak ties in sharing knowledge across organization subunits. \emph{Admin. Sci. Quart.} 44:82-111.

Hargadon A. 2002. Brokering knowledge: Linking learning and innovation. \emph{Res. Organ. Behavior} 24:41-86.

Hargadon A, Sutton RI. 1997. Technology brokering and innovation in a product development firm. \emph{Admin. Sci. Quart.} 42:716-749.

Harhoff D, Narin F, Scherer FM, Vopel K. 1999. Citation frequency and the value of patented inventions. \emph{Rev. Econom. Statist.} 81:511-515.

Hausman J, Hall BH, Griliches Z. 1984. Econometric models for count data with an application to the patents-R\&D relationship. \emph{Econometrica} 52:909-938.

Hirsch JE. 2005. An Index to Quantify an Individual's Scientific Research Output. \emph{Proc. Nat. Acad. Sci.}  102:16569-16572.

Hughes SS. 2001. Making dollars out of DNA: The first major patent in biotechnology and the commercialization of molecular biology, 1974-1980. \emph{Isis} 92:541-575.

Iacus SM, King G, Porro G. 2011. Multivariate matching methods that are monotonic imbalance bounding. \emph{J. Amer. Statist. Assoc.} 106:345-361.

Jaffe AB, Lerner J. 2004. \emph{Innovation and Its Discontents}. Princeton, NJ: Princeton University Press.

Jaffe AB, Trajtenberg M, Fogarty MS. 2002. The meaning of patent citations: Report on the NBER/Case-Western Reserve survey of patentees. In \emph{Patents, Citations \& Innovations}, edited by AB Jaffe and M Trajtenberg. Cambridge: MIT Press.

Kenney M, Patton D. 2009. Reconsidering the Bayh-Dole act and the current university invention ownership model. \emph{Res. Policy} 38:1407-1422.

Klepper S. 1997. Industry life cycles. \emph{Indust. Corporate Change} 6:145-182.

Krimsky S. 2004. \emph{Science In the Private Interest}. Lanham: Rowman \& Littlefield Publishers.

Lai R, D'Amour A, Yu A, Sun Y, Torvik V, Fleming L. 2011. Disambiguation and co-authorship networks of the US patent inventor database. Cambridge, MA: Harvard Business School.

Lane J. 2010. Let's make science metrics more scientific. \emph{Nature} 464:488-489.

Lanjouw JO, Schankerman M. 2004. Patent quality and research productivity: Measuring innovation with multiple indicators. \emph{Econom. J.} 114:441-465.

Lemley MA, Sampat B. 2008. Is the patent office a rubber stamp? \emph{Emory Law J.} 58:181-206.

Levy S. 2011. \emph{In the Plex: How Google Thinks, Works, and Shapes Our Lives}. New York: Simon \& Schuster.

Lucas HC, Goh JM. 2009. Disruptive technology: How kodak missed the digital photography revolution. \emph{J. Strategic Information Systems} 18:46-55.

Mody CCM. 2011. \emph{Instrumental Community}. Cambridge, MA: MIT Press.

Mowery, DC, Sampat BN, Ziedonis AA. 2002. Learning to patent: Institutional experience, learning, and the characteristics of US university patents after the Bayh-Dole act, 1981-1992. \emph{Management Sci.} 48:73-89.

Murmann JP. 2003. \emph{Knowledge and Competitive Advantage} Cambridge: Cambridge University Press.

Murray F. 2002. Innovation as co-evolution of scientific and technological networks: Exploring tissue engineering. \emph{Res. Policy} 31:1389-1403.

Nelson RR, Winter SG. 1982. \emph{An Evolutionary Theory of Economic Change}. Cambridge, MA: Belknap Press of Harvard University Press.

Orr D. 1974. Index of entry barriers and its application to market structure performance relationship. \emph{J. Indust. Econom.} 23:39-49.

Owen-Smith J, Powell WW. 2003. The expanding role of university patenting in the life sciences: Assessing the importance of experience and connectivity. \emph{Res. Policy} 32:1695-1711.

Owen-Smith J, Powell WW. 2004. Knowledge networks as channels and conduits: The effects of spillovers in the boston biotechnology community. \emph{Organ. Sci.} 15:5-21.

Pollack A. 2009. As patent ends, a seed's use will survive. \emph{New York Times} December 18, 2009.

Powell WW, Koput KW, Smith-Doerr L. 1996. Interorganizational collaboration and the locus of innovation: Networks of learning in biotechnology. \emph{Admin. Sci. Quart.} 41:116-145.

Powell WW, Owen-Smith J. 1998. Universities and the market for intellectual property in the life sciences. \emph{J. Policy Anal. Management} 17:253-277.

Powell WW, Owen-Smith J, Colyvas J. 2007. Innovaton and emulation: Lessons from American universities in selling private rights to public knowledge. \emph{Minerva} 45:121-142.

Powell WW, White DR, Koput KW, Owen-Smith J. 2005. Network dynamics and field evolution: The growth of interorganizational collaboration in the life sciences. \emph{Amer. J. Sociol.} 110:1132-1205.

Price DJdS. 1965. Networks of scientific papers. \emph{Sci} 149:510-515.

Rao H. 2009. \emph{Market Rebels} Princeton: Princeton University Press.

Rossman G, Chiu MM, Mol JM. 2008. Modeling diffusion of multiple innovations via multilevel diffusion curves: Payola in pop music radio. \emph{Sociol. Methodology} 38:201-230.

Schumpeter J. 1942. \emph{Capitalism, Socialism and Democracy}. New York: Perennial.

Sine WD, Shane S, Gregorio DD. 2003. The halo effect and technology licensing: The influence of institutional prestige on the licensing of university inventions. \emph{Management Sci.} 49:478-496.

Singh J, Agrawal A. 2011. Recruiting for Ideas: How Firms Exploit the Prior Inventions of New Hires. \emph{Management Sci.} 57:129-150.

Singh J, Fleming L. 2010. Lone inventors as sources of breakthroughs: Myth or reality? \emph{Management Sci.} 56:41-56.

Sosa ML. 2009. Application-Specific R\&D Capabilities and the Advantage of Incumbents: Evidence from the Anticancer Drug Market. \emph{Management Sci.} 55:1409-1422.

Sosa ML. 2011. From old competence destruction to new competence access: Evidence from the comparison of two discontinuities in anticancer drug discovery. \emph{Organ. Sci.} 22:1500-1516.

Sosa ML, Murray FE, Utterback JM. 2012. Competence-enhancing discontinuities as capability-adding transitions: The contrasting effects of biotechnology and nanotechnology on anticancer drug discovery. London: London Business School.

Strang D, Macy MW. 2001. In search of excellence: Fads, success stories, and adaptive emulation. \emph{Amer. J. Sociol.} 107:147-182.

Stuart T, Sorenson O. 2003. The geography of opportunity: Spatial heterogeneity in founding rates and the performance of biotechnology firms. \emph{Res. Policy} 32:229-253.

Teece DJ. 1986. Profiting from technological innovation: Implications for integration, collaboration, licensing and public policy. \emph{Res. Policy} 15:285-305.

Trajtenberg M, Henderson R, Jaffe AB. 1997. University versus corporate patents: A window on the basicness of invention. \emph{Econom. Innovation New Tech.} 5:19-50.

Tushman ML, Anderson P. 1986. Technological discontinuities and organizational environments. \emph{Admin. Sci. Quart.} 31:439-465.

Washburn J. 2005. \emph{University, Inc.} New York: Basic Books.

Whittington KB, Owen-Smith J, Powell WW. 2009. Networks, propinquity, and innovation in knowledge-intensive industries. \emph{Admin. Sci. Quart.} 54:90-122.

Wuchty S, Jones BF, Uzzi B. 2007. The increasing dominance of teams in production of knowledge. \emph{Sci.} 316:1036-1039.

Youtie J, Iacopette M, Graham S. 2008. Assessing the nature of nanotechnology: Can we uncover an emerging general purpose technology? \emph{J. Tech. Transfer} 33:315-329.

\end{hangparas}

\pagebreak

\tikzset{
	node/.style={circle,inner sep=0.5mm,minimum size=0.3cm,draw,thin,black,fill=black,text=black},
	nondirectional/.style={thin,black}
}

\begin{sidewaysfigure}
\centering
\begin{tikzpicture}[scale=1, font=\sffamily]

\node [] (net) at (-1.0, 0)  {\setlength\fboxsep{0pt}
\setlength\fboxrule{0.5pt}
\fbox{\includegraphics[scale=0.45]{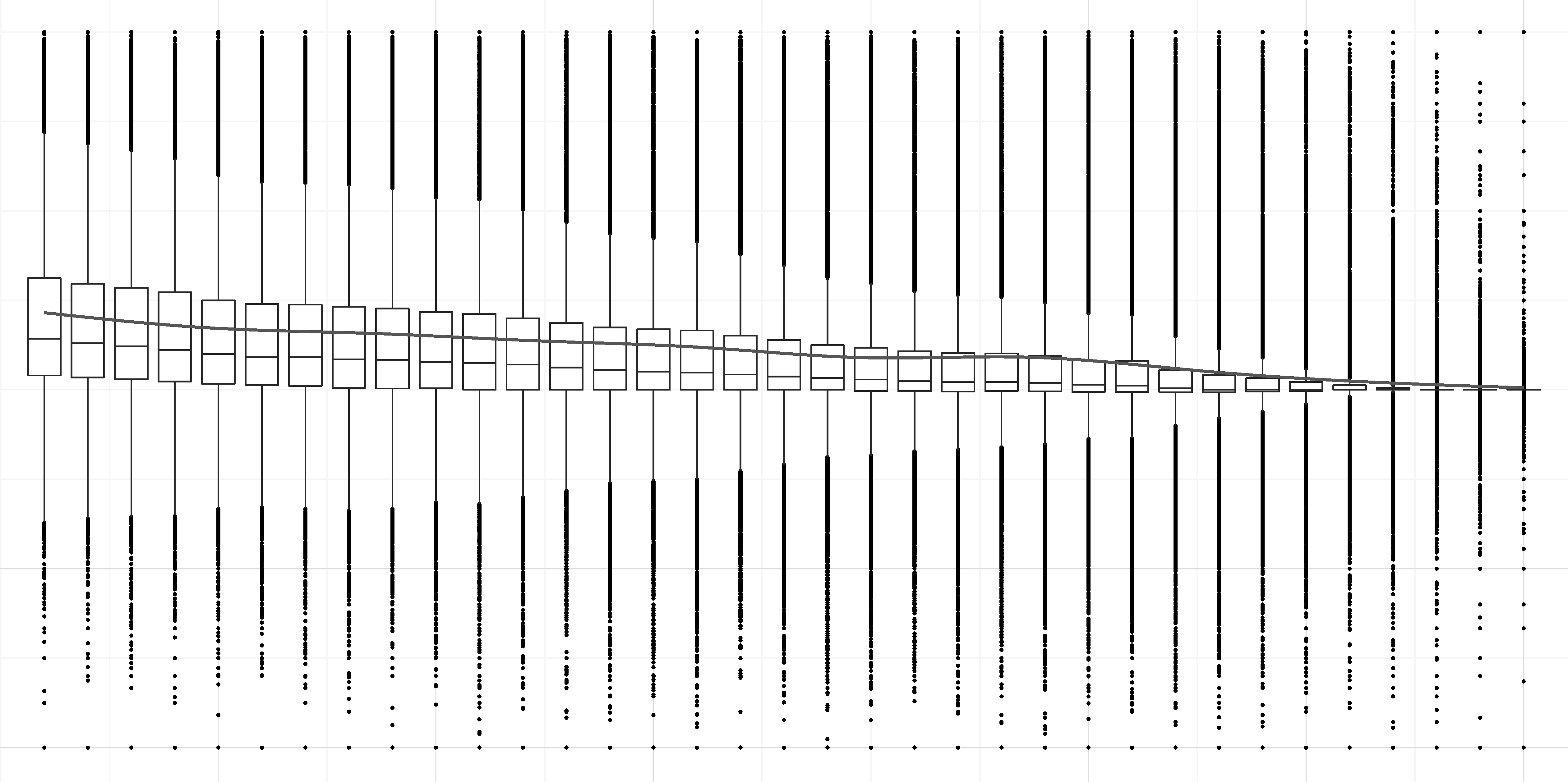}}};

			\node [rotate=90] (i1) at (-12, 0.000)	{{\large Disruptiveness}};

\draw (-9.97,-5.03) node[fill=white] {\tiny 1976};
\draw (-7.86,-5.03) node[fill=white] {\tiny 1980};

\draw (-5.75,-5.03) node[fill=white] {\tiny 1984};

\draw (-3.59,-5.03) node[fill=white] {\tiny 1988};

\draw (-1.45,-5.03) node[fill=white] {\tiny 1992};

\draw (0.67,-5.03) node[fill=white] {\tiny 1996};

\draw (2.86,-5.03) node[fill=white] {\tiny 2000};

\draw (5.0,-5.03) node[fill=white] {\tiny 2004};

\draw (7.1,-5.03) node[fill=white] {\tiny 2008};

\draw (-11.2,4.40) node[fill=white] {\tiny 1.0};
\draw (-11.2,2.21) node[fill=white] {\tiny 0.5};
\draw (-11.2,0) node[fill=white] {\tiny 0.0};
\draw (-11.2,-2.18) node[fill=white] {\tiny -0.5};

\end{tikzpicture}

\caption{\textsc{Fig. 1.---}Distribution of disruptiveness among all U.S. utility patents (1976-2010).}

\end{sidewaysfigure}
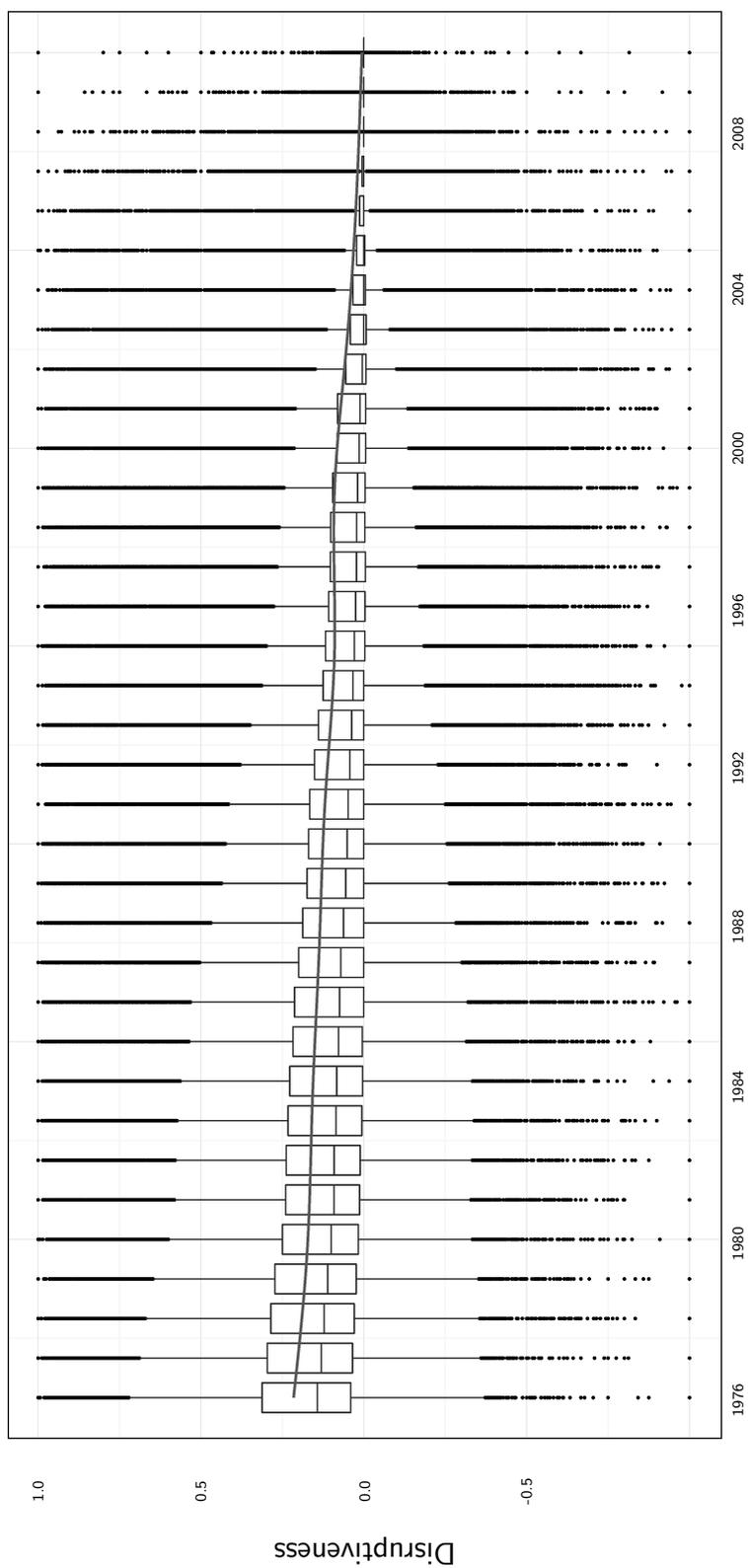

\pagebreak

\tikzset{
	node/.style={rectangle,inner sep=0.5mm,minimum size=0.2cm,draw,thin,black,fill=black,text=black},
	unidirectional/.style={thin,shorten <= 0pt, shorten >= 2pt, -stealth}
}

\begin{figure}[ht!]

\centering
 \includegraphics[angle=270,width=12.5cm]{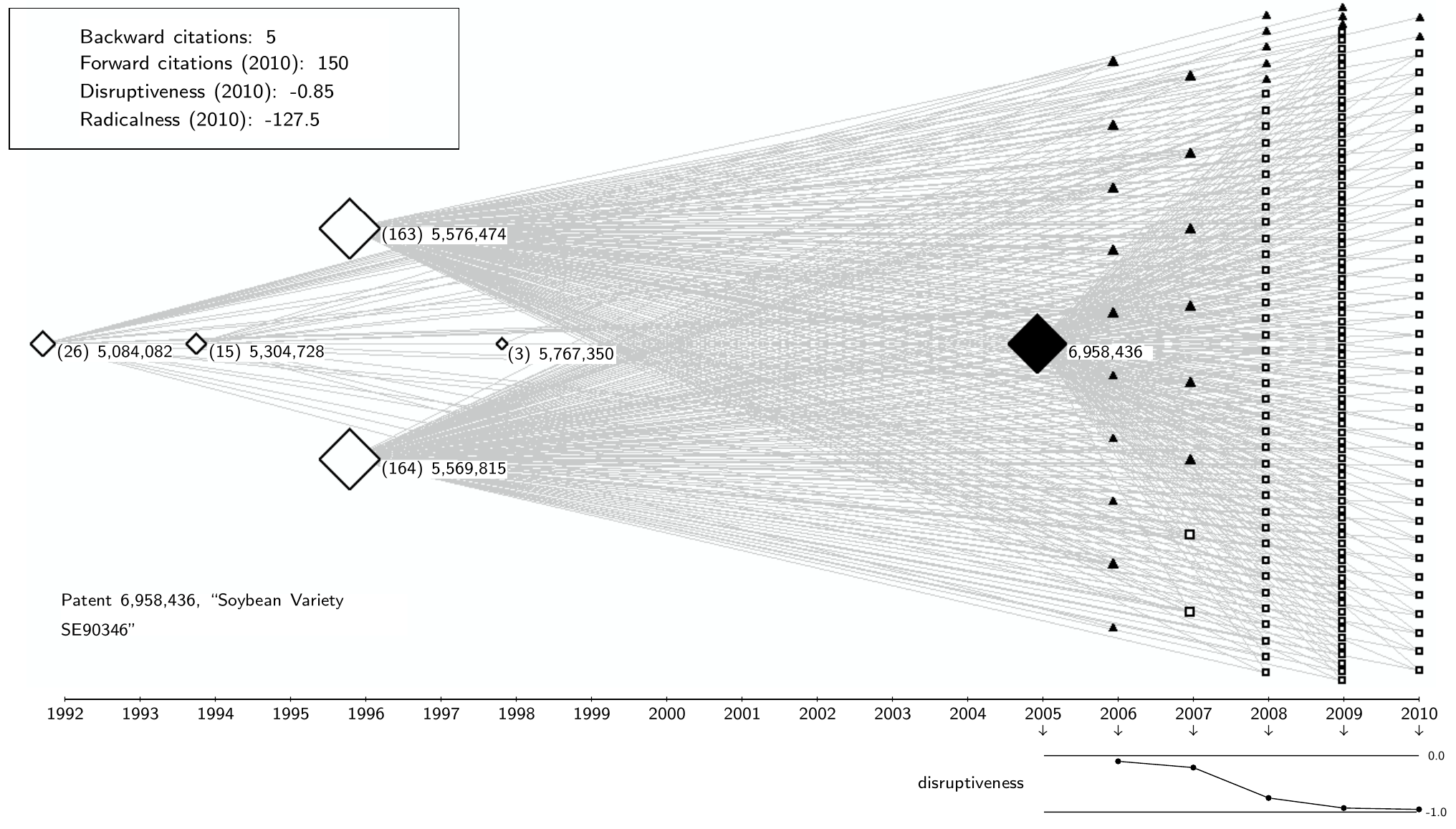}

 \includegraphics[angle=270,width=12.5cm]{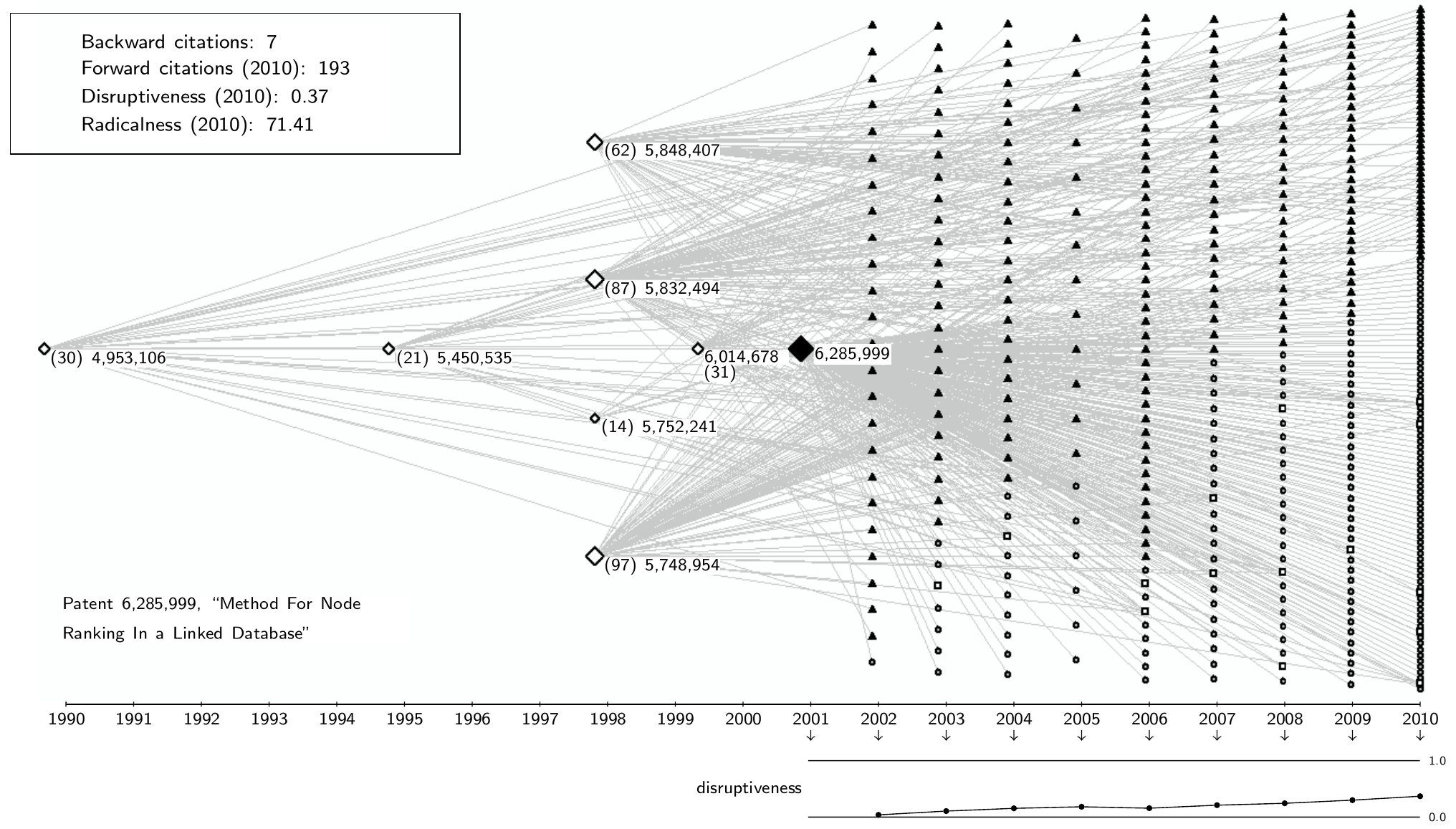}

 \includegraphics[angle=270,width=12.5cm]{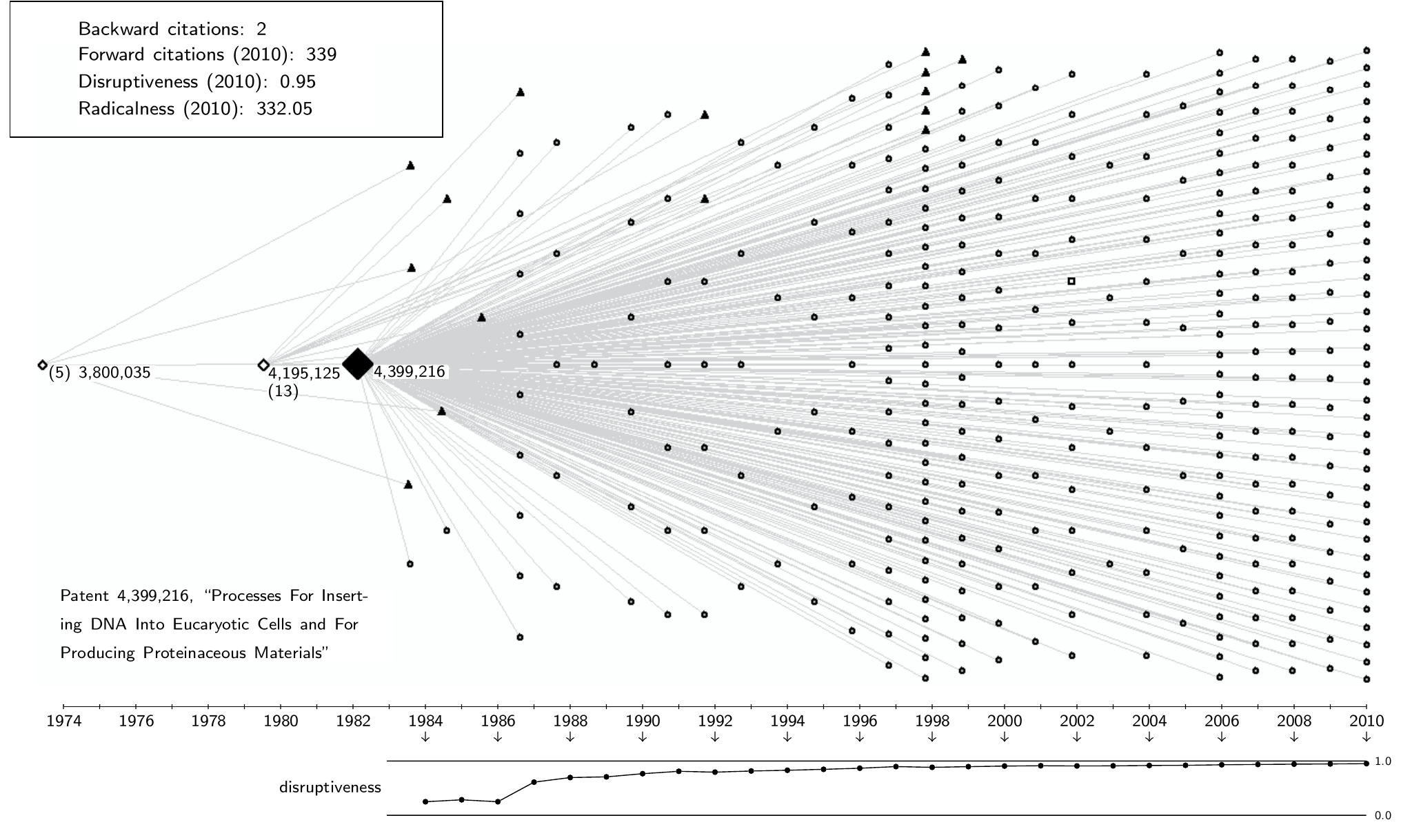}

 \caption{\small \textsc{Fig. 2.---}Network diagrams for the Monsanto, PageRank, and Axel patents. The focal patent is a black diamond; its prior art is indicated by white diamonds.  Triangles cite the prior art but not the focal patent, squares cite the focal patent and its prior art, and circles cite the focal patent but not the prior art.  The degree centralities of the focal patent prior art are given in parentheses.  Node sizes of the focal patent and prior art are proportional to the degree centralities. Disruptiveness over time is shown below each network.}
 
 \end{figure}

\pagebreak

\begin{sidewaysfigure}
\centering

\begin{tikzpicture}[scale=0.9]
    \begin{axis}[
        title=Mean Citations Per Year Relative to Grant Date of Focal Patent,
        xlabel=Time to and From Focal Patent Grant (Years),
        ylabel=Mean Citations,
        ymax=1.3,
        ymin=0.45,
        xmin=-5,
        xmax=10,
        yscale=1,
        xscale=1,
        width=22cm,
        height=15cm,
        enlargelimits=false,
        legend style={at={(0.04,0.95)},anchor=north west},
        every axis legend/.append style={nodes={right}}]
   
        \addplot[color=black, very thick, dashed]
        plot coordinates {
(	-10	,	0.2711	)
(	-9	,	0.324	)
(	-8	,	0.4335	)
(	-7	,	0.4374	)
(	-6	,	0.5651	)
(	-5	,	0.5296	)
(	-4	,	0.5773	)
(	-3	,	0.5969	)
(	-2	,	0.6709	)
(	-1	,	0.8739	)
(	0	,	1.0269	)
(	1	,	1.146	)
(	2	,	1.2429	)
(	3	,	1.1661	)
(	4	,	1.1027	)
(	5	,	1.0939	)
(	6	,	1.0313	)
(	7	,	1.0362	)
(	8	,	1.0441	)
(	9	,	0.9612	)
(	10	,	0.902	)
(	11	,	0.9124	)
(	12	,	0.9281	)
(	13	,	0.8024	)
(	14	,	0.8959	)
(	15	,	0.6688	)
(	16	,	0.581	)
(	17	,	0.6702	)
(	18	,	0.5986	)
(	19	,	0.6035	)
(	20	,	0.5588	)
(	21	,	0.6273	)
(	22	,	0.5991	)
(	23	,	0.4933	)
(	24	,	0.3396	)
(	25	,	0.4868	)
(	26	,	0.3209	)
(	27	,	0.3038	)
(	28	,	0.3828	)
(	29	,	0.1649	)
(	30	,	0.3016	)				};
    \addlegendentry{Control}

   \addplot[color=black, densely dotted, very thick]
        plot coordinates {
(	-10	,	0.268	)
(	-9	,	0.296	)
(	-8	,	0.3707	)
(	-7	,	0.4716	)
(	-6	,	0.4743	)
(	-5	,	0.5217	)
(	-4	,	0.5398	)
(	-3	,	0.5983	)
(	-2	,	0.6235	)
(	-1	,	0.783	)
(	0	,	0.9538	)
(	1	,	0.9006	)
(	2	,	0.8769	)
(	3	,	0.8201	)
(	4	,	0.787	)
(	5	,	0.706	)
(	6	,	0.6864	)
(	7	,	0.6209	)
(	8	,	0.6453	)
(	9	,	0.5717	)
(	10	,	0.5701	)
(	11	,	0.5475	)
(	12	,	0.5478	)
(	13	,	0.509	)
(	14	,	0.4639	)
(	15	,	0.4379	)
(	16	,	0.4377	)
(	17	,	0.3822	)
(	18	,	0.3616	)
(	19	,	0.4093	)
(	20	,	0.3804	)
(	21	,	0.353	)
(	22	,	0.2815	)
(	23	,	0.32	)
(	24	,	0.3333	)
(	25	,	0.3132	)
(	26	,	0.3535	)
(	27	,	0.4557	)
(	28	,	0.4688	)
(	29	,	0.3608	)
(	30	,	0.4921	)
	};
     \addlegendentry{Treated}

   \addplot[color=black, solid, very thick]
        plot coordinates {
(	-0	,	0	)
(	-0	,	1.5)
	};

 \addplot graphics[xmin=0.13,ymin=.475,xmax=8.7,ymax=0.595] {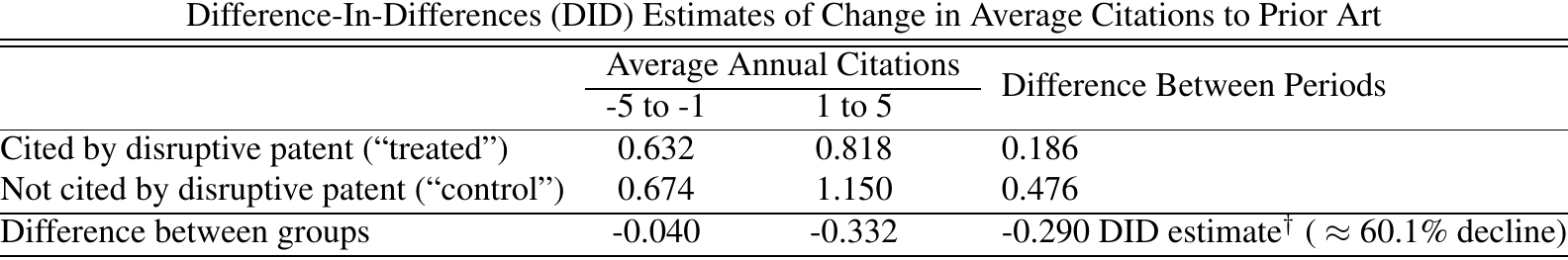};
 
    \end{axis}

    \end{tikzpicture}

\caption{\textsc{Fig. 3.---}Lines correspond to the mean citations to focal patent prior art relative to the grant date of the focal patent.  Total $N=119,486$ focal patent--prior art--year pairs are represented in the figure, or 59,743 ``treated'' and 59,743 ``control'' pairs. Significance of the DID estimate was evaluated with a regression on the disaggregated data: $t=-7.772$, $df=5,491$, and $p<0.001$ (two-tailed test).  Standard errors were estimated using a block bootstrap to correct for serial correlation (see Bertrand, Duflo, \& Mullainathan, [2004]).}

\label{figITSraw}
\end{sidewaysfigure}
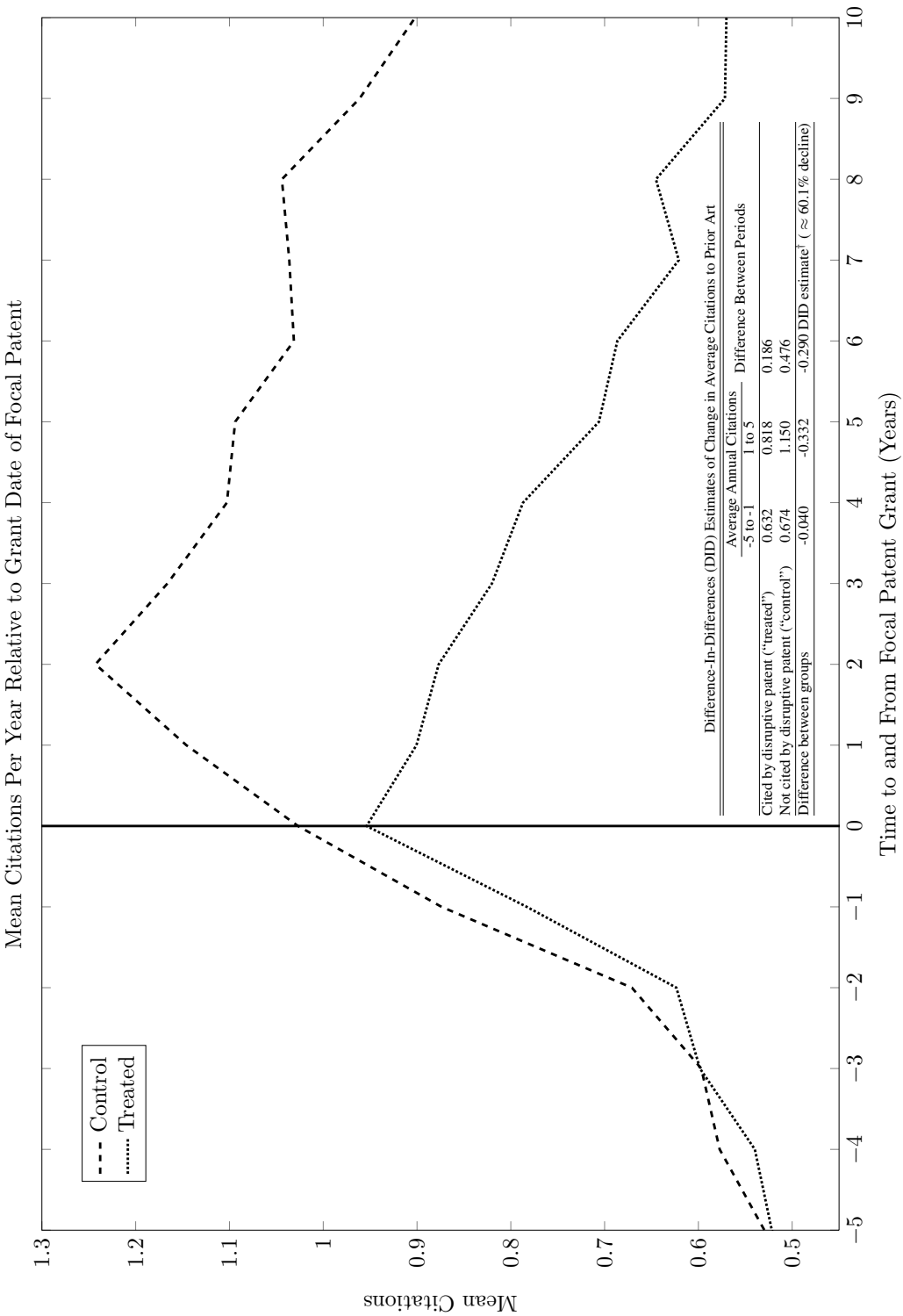


\pagebreak

\clearpage
\begin{sidewaystable}[h!]\centering
\def\sym#1{\ifmmode^{#1}\else\(^{#1}\)\fi}

\caption{Table 1\\Descriptive Statistics and Correlations, Top 0.25\% Most Cited Utility Patents (1976-2010)$^\dagger$}
\scalebox{0.9}{
\small\addtolength{\tabcolsep}{-5pt}
\renewcommand{\arraystretch}{1}
\begin{tabular}{l*{13}{D{.}{.}{-1}}}
\hline\hline																							
\multicolumn{1}{c}{Variable}& \multicolumn{1}{c}{Mean} & \multicolumn{1}{c}{SD} & \multicolumn{1}{c}{Min} & \multicolumn{1}{c}{Max}	&	1	&	2	&	3	&	4	&	5	&	6	&	7	&	8	&	9		\\
\hline																							
1.	Forward citations (impact)	&	212.67	&	98.37	&	143	&	2211	&	1.00	&		&		&		&		&		&		&		&		\\
2.	Backward citations	&	17.22	&	27.51	&	0	&	656	&	0.02	&	1.00	&		&		&		&		&		&		&		\\
3.	Disruptiveness	&	0.09	&	0.31	&	-0.90	&	1.00	&	0.05	&	-0.20	&	1.00	&		&		&		&		&		&		\\
4.	Forward citations to focal patent only	&	115.31	&	97.09	&	0	&	2211	&	0.68	&	-0.19	&	0.62	&	1.00	&		&		&		&		&		\\
5.	Forward citations to prior art only	&	782.72	&	1004.85	&	0	&	16226	&	0.09	&	0.76	&	-0.24	&	-0.17	&	1.00	&		&		&		&		\\
6.	Forward citations to focal patent and prior art	&	97.36	&	78.58	&	0	&	794	&	0.42	&	0.26	&	-0.71	&	-0.39	&	0.33	&	1.00	&		&		&		\\
7.	Forward citations to prior art before focal patent grant	&	215.39	&	451.76	&	0	&	12741	&	0.01	&	0.87	&	-0.15	&	-0.14	&	0.81	&	0.19	&	1.00	&		&		\\
8.	Forward citations to prior art year before focal patent grant	&	73.23	&	200.83	&	0	&	4095	&	0.01	&	0.82	&	-0.13	&	-0.15	&	0.75	&	0.19	&	0.90	&	1.00	&		\\
9.	Application year	&	1991.46	&	5.80	&	1957	&	2004	&	-0.07	&	0.22	&	-0.21	&	-0.19	&	0.30	&	0.15	&	0.30	&	0.25	&	1.00	\\
10.	Grant year	&	1993.69	&	5.86	&	1976	&	2006	&	-0.06	&	0.24	&	-0.21	&	-0.18	&	0.32	&	0.14	&	0.32	&	0.28	&	0.98	\\
11.	Scientific references	&	6.32	&	20.94	&	0	&	823	&	0.05	&	0.48	&	-0.04	&	-0.02	&	0.42	&	0.08	&	0.50	&	0.43	&	0.15	\\
12.	Claims	&	24.34	&	23.46	&	1	&	868	&	0.04	&	0.18	&	-0.04	&	0.00	&	0.17	&	0.05	&	0.15	&	0.13	&	0.11	\\
13.	Inventor team size	&	2.62	&	1.99	&	1	&	34	&	0.01	&	0.08	&	0.02	&	0.05	&	0.06	&	-0.05	&	0.09	&	0.08	&	0.14	\\
14.	Government interest (1 = yes)	&	0.02	&	0.15	&	0	&	1	&	-0.01	&	-0.02	&	0.05	&	0.03	&	-0.03	&	-0.04	&	-0.02	&	-0.02	&	-0.02	\\
15.	Total assignees	&	0.88	&	0.35	&	0	&	3	&	-0.02	&	0.05	&	0.04	&	0.01	&	0.03	&	-0.03	&	0.05	&	0.06	&	0.13	\\
16.	Unassigned (assignee 1)	&	0.13	&	0.33	&	0	&	1	&	0.01	&	-0.05	&	-0.04	&	-0.01	&	-0.03	&	0.03	&	-0.05	&	-0.05	&	-0.12	\\
17.	Individual (assignee 1	&	0.00	&	0.04	&	0	&	1	&	0.01	&	0.00	&	-0.01	&	0.01	&	0.02	&	0.00	&	0.01	&	0.01	&	0.02	\\
18.	Firm (assignee 1)	&	0.82	&	0.38	&	0	&	1	&	-0.01	&	0.06	&	-0.01	&	-0.01	&	0.06	&	0.00	&	0.07	&	0.07	&	0.13	\\
19.	Government (assignee 1)	&	0.01	&	0.09	&	0	&	1	&	-0.02	&	-0.03	&	0.06	&	0.02	&	-0.05	&	-0.05	&	-0.03	&	-0.03	&	-0.06	\\
20.	Nonprofit (assignee 1)	&	0.01	&	0.08	&	0	&	1	&	0.00	&	-0.02	&	0.02	&	0.00	&	0.00	&	-0.01	&	-0.02	&	-0.02	&	-0.04	\\
21.	University (assignee 1)	&	0.04	&	0.19	&	0	&	1	&	0.01	&	-0.03	&	0.05	&	0.02	&	-0.04	&	-0.02	&	-0.03	&	-0.03	&	-0.02	\\
	\hline
\multicolumn{1}{c}{Variable}	&	10 & 11	&	12	&	13	&	14	&	15	&	16	&	17	&	18	&	19	& 20 & 21 	\\
\hline																							
10.	Grant year	&	1.00	&		&		&		&		&		&		&		&		&		&		&		\\
11.	Scientific references	&	0.17	&	1.00	&		&		&		&		&		&		&		&		&		&		\\
12.	Claims	&	0.13	&	0.13	&	1.00	&		&		&		&		&		&		&		&		&		\\
13.	Inventor team size	&	0.15	&	0.07	&	0.10	&	1.00	&		&		&		&		&		&		&		&		\\
14.	Government interest (1 = yes)	&	-0.02	&	0.04	&	0.00	&	0.02	&	1.00	&		&		&		&		&		&		&		\\
15.	Total assignees	&	0.14	&	0.03	&	0.05	&	0.21	&	0.04	&	1.00	&		&		&		&		&		&		\\
16.	Unassigned (assignee 1)	&	-0.12	&	-0.03	&	-0.05	&	-0.21	&	-0.04	&	-0.97	&	1.00	&		&		&		&		&		\\
17.	Individual (assignee 1	&	0.03	&	0.01	&	0.01	&	-0.01	&	-0.01	&	0.03	&	-0.01	&	1.00	&		&		&		&		\\
18.	Firm (assignee 1)	&	0.13	&	-0.01	&	0.03	&	0.18	&	-0.21	&	0.77	&	-0.81	&	-0.08	&	1.00	&		&		&		\\
19.	Government (assignee 1)	&	-0.06	&	-0.01	&	-0.02	&	-0.01	&	0.15	&	0.03	&	-0.03	&	0.00	&	-0.18	&	1.00	&		&		\\
20.	Nonprofit (assignee 1)	&	-0.04	&	0.00	&	-0.01	&	-0.01	&	0.05	&	0.03	&	-0.03	&	0.00	&	-0.17	&	-0.01	&	1.00	&		\\
21.	University (assignee 1)	&	-0.01	&	0.07	&	0.03	&	0.02	&	0.39	&	0.09	&	-0.08	&	-0.01	&	-0.43	&	-0.02	&	-0.02	&	1.00	\\
\hline\hline																			$^\dagger N = 9,747$	
\end{tabular}																							
}																							
\label{tableDescriptives}																							
\end{sidewaystable}																																																				

\pagebreak
\clearpage
\begin{sidewaystable}[h!]\centering
\def\sym#1{\ifmmode^{#1}\else\(^{#1}\)\fi}

\caption{Table 2\\Illustrative Patents}
\scalebox{0.85}{
\small\addtolength{\tabcolsep}{-5pt}
\renewcommand{\arraystretch}{1}
\begin{tabular}{lcccccccccc}
\hline\hline																							
Patent	&	\multicolumn{1}{m{1.2cm}}{\centering Forward cites}	&	\multicolumn{1}{m{1.5cm}}{\centering Backward cites}	&	Disruptiveness	&	\multicolumn{1}{m{1.5cm}}{\centering Forward cites to focal only}	&	\multicolumn{1}{m{1.5cm}}{\centering Forward cites to prior only}	&	\multicolumn{1}{m{2cm}}{\centering Forward cites to focal and prior}	&	\multicolumn{1}{m{1.5cm}}{\centering Application year}	&	\multicolumn{1}{m{1.5cm}}{\centering Grant year}	&	Title	&	Assignee	\\
\hline																							

4,637,464	&	194	&	7	&	-0.90	&	2	&	17	&	192	&	1984	&	1987	&	\multicolumn{1}{m{5cm}}{In situ retorting of oil shale with pulsed water purge}	&	\multicolumn{1}{m{4.5cm}}{\hspace{3 mm}Amoco Corp.}	\\
4,573,530	&	192	&	6	&	-0.89	&	1	&	21	&	191	&	1983	&	1986	&	\multicolumn{1}{m{5cm}}{In-situ gasification of tar sands utilizing a combustible gas}	&	\multicolumn{1}{m{4.5cm}}{\hspace{3 mm}Mobil Oil Corp.}	\\
4,658,215	&	200	&	4	&	-0.87	&	7	&	13	&	193	&	1986	&	1987	&	\multicolumn{1}{m{5cm}}{Method for induced polarization logging}	&	\multicolumn{1}{m{4.5cm}}{\hspace{3 mm}Shell Oil Co.}	\\
4,928,765	&	195	&	10	&	-0.85	&	2	&	29	&	193	&	1988	&	1990	&	\multicolumn{1}{m{5cm}}{Method and apparatus for shale gas recovery}	&	\multicolumn{1}{m{4.5cm}}{\hspace{3 mm}Ramex Synfuels, Int'l, Inc.}	\\
6,958,436	&	150	&	5	&	-0.85	&	0	&	26	&	150	&	2002	&	2005	&	\multicolumn{1}{m{5cm}}{Soybean variety SE90346}	&	\multicolumn{1}{m{4.5cm}}{\hspace{3 mm}Monsanto Co.}	\\
5,015,744	&	173	&	4	&	-0.36	&	32	&	129	&	141	&	1989	&	1991	&	\multicolumn{1}{m{5cm}}{Method for preparation of taxol using an oxazinone}	&	\multicolumn{1}{m{4.5cm}}{\hspace{3 mm}Florida State University}	\\
6,376,284	&	175	&	19	&	-0.24	&	14	&	446	&	161	&	2000	&	2002	&	\multicolumn{1}{m{5cm}}{Method of fabricating a memory device}	&	\multicolumn{1}{m{4.5cm}}{\hspace{3 mm}Micron Technology, Inc.}	\\
6,063,738	&	178	&	12	&	-0.14	&	65	&	161	&	113	&	1999	&	2000	&	\multicolumn{1}{m{5cm}}{Foamed well cement slurries, additives and methods}	&	\multicolumn{1}{m{4.5cm}}{\hspace{3 mm}Halliburton Co.}	\\
4,724,318	&	145	&	2	&	0.12	&	89	&	132	&	56	&	1986	&	1988	&	\multicolumn{1}{m{5cm}}{Atomic force microscope and method for imaging surfaces with atomic resolution}	&	\multicolumn{1}{m{4.5cm}}{\hspace{3 mm}IBM Corp.}	\\
5,016,107	&	163	&	17	&	0.14	&	126	&	482	&	37	&	1989	&	1991	&	\multicolumn{1}{m{5cm}}{Electronic still camera utilizing image compression and digital storage}	&	\multicolumn{1}{m{4.5cm}}{\hspace{3 mm}Eastman Kodak Co.	}\\
6,285,999	&	193	&	7	&	0.37	&	178	&	248	&	15	&	1998	&	2001	&	\multicolumn{1}{m{5cm}}{Method for node ranking in a linked database}	&	\multicolumn{1}{m{4.5cm}}{\hspace{3 mm}Stanford University}	\\
4,356,429	&	409	&	4	&	0.66&	358	&	358	&	51	&	1980	&	1982	&	\multicolumn{1}{m{5cm}}{Organic electroluminescent cell}	&	\multicolumn{1}{m{4.5cm}}{\hspace{3 mm}Eastman Kodak Co.}	\\
4,445,050	&	151	&	4	&	0.89	&	151	&	18	&	0	&	1981	&	1984	&	\multicolumn{1}{m{5cm}}{Device for conversion of light power to electric power}	&	\multicolumn{1}{m{4.5cm}}{\hspace{3 mm}None}	\\
5,010,405	&	159	&	2	&	0.92&	159	&	14	&	0	&	1989	&	1991	&	\multicolumn{1}{m{5cm}}{Receiver-compatible enhanced definition television system}	&	\multicolumn{1}{m{4.5cm}}{\hspace{3 mm}MIT}	\\
4,237,224	&	282	&	1	&	0.94	&	277	&	8	&	5	&	1979	&	1980	&	\multicolumn{1}{m{5cm}}{Process for producing biologically functional molecular chimeras}	&	\multicolumn{1}{m{4.5cm}}{\hspace{3 mm}Stanford University}	\\
4,399,216	&	339	&	2	&	0.95	&	338	&	15	&	1	&	1980	&	1983	&	\multicolumn{1}{m{5cm}}{Processes for inserting DNA into eucaryotic cells and for producing proteinaceous materials}	&	\multicolumn{1}{m{4.5cm}}{\hspace{3 mm}Columbia University}	\\
4,343,993	&	169	&	0	&	1.00	&	169	&	0	&	0	&	1980	&	1982	&	\multicolumn{1}{m{5cm}}{Scanning tunneling microscope}	&	\multicolumn{1}{m{4.5cm}}{\hspace{3 mm}IBM Corp.}	\\
4,683,202	&	2,211	&	0	&	1.00	&	2,211	&	0	&	0	&	1985	&	1987	&	\multicolumn{1}{m{5cm}}{Process for amplifying nucleic acid sequences}	&	\multicolumn{1}{m{4.5cm}}{\hspace{3 mm}Cetus Corp.}	\\
\hline\hline																					
\end{tabular}																							
}																							
\label{tableDescriptives}																							
\end{sidewaystable}

\pagebreak
\clearpage
\begin{sidewaystable}[h!]\centering
\def\sym#1{\ifmmode^{#1}\else\(^{#1}\)\fi}

\caption{Table 3\\University Variable Descriptive Statistics and Correlations$^\dagger$}
\scalebox{1.0}{
\small\addtolength{\tabcolsep}{-5pt}
\renewcommand{\arraystretch}{1}
\begin{tabular}{l*{9}{D{.}{.}{-1}}}
\hline\hline																							
\multicolumn{1}{c}{Variable}	&	\multicolumn{1}{c}{Mean}	&	\multicolumn{1}{c}{SD}	&	\multicolumn{1}{c}{Min}	&	\multicolumn{1}{c}{Max}	&	1	&	2	&	3	&	4	&	5	\\
\hline																							
1. Radicalness	&	40.26	&	78.64	&	-35.11	&	764.08	&	1.00	&		&		&		&		\\
2. Disruptiveness	&	0.10	&	0.11	&	-0.71	&	1.00	&	0.40	&	1.00	&		&		&		\\
3. New combinations	&	4.28	&	6.68	&	0.00	&	76.93	&	-0.09	&	-0.18	&	1.00	&		&		\\
4. Originality	&	0.40	&	0.12	&	0.00	&	0.88	&	-0.17	&	-0.50	&	0.05	&	1.00	&		\\
5. Impact	&	303.86	&	494.75	&	0&	4546	&	0.86	&	0.16	&	-0.09	&	0.01	&	1.00	\\
6. Volume (patents)	&	29.31	&	27.69	&	1	&	186	&	0.50	&	-0.03	&	0.08	&	0.08	&	0.67	\\
7. Technology transfer age (decades)	&	1.55	&	1.18	&	0.00	&	7.90	&	0.16	&	-0.11	&	0.16	&	0.06	&	0.28	\\
8. Life sciences dominant	&	0.49	&	0.50	&	0	&	1&	-0.01	&	0.13	&	0.03	&	-0.31	&	-0.10	\\
9. Scientific articles (log)	&	7.43	&	0.59	&	5.72	&	9.08	&	0.32	&	0.03	&	0.09	&	-0.01	&	0.35	\\
10. Impact factor	&	1.68	&	0.53	&	0.69	&	4.89	&	0.28	&	0.07	&	0.10	&	-0.10	&	0.25	\\
11. Industry sponsored R\&D (millions)	&	14.28	&	14.89	&	0.07	&	122.18	&	0.13	&	-0.12	&	0.07	&	0.14	&	0.27	\\
12. Industry contractual ties	&	1.37	&	2.31	&	0&	22	&	0.27	&	0.01	&	0.05	&	0.02	&	0.32	\\
13. NSF grants (log)	&	9.41	&	1.27	&	5.06	&	11.82	&	0.17	&	-0.09	&	0.08	&	0.17	&	0.26	\\
14. NIH grants (log)	&	10.74	&	1.21	&	6.21	&	13.30	&	0.18	&	-0.01	&	0.16	&	-0.07	&	0.16	\\\hline																							
\multicolumn{1}{c}{Variable}	&	7 & 8	&	9	&	10	&	11	&	12	&	13	&	14		\\
\hline																							
6. Volume (patents)	&	1.00	&		&		&		&		&		&		&		&		\\
7. Technology transfer age (decades)	&	0.48	&	1.00	&		&		&		&		&		&		&		\\
8. Life sciences dominant	&	-0.03	&	0.00	&	1.00	&		&		&		&		&		&		\\
9. Scientific articles (log)	&	0.62	&	0.33	&	0.12	&	1.00	&		&		&		&		&		\\
10. Impact factor	&	0.37	&	0.09	&	0.32	&	0.29	&	1.00	&		&		&		&		\\
11. Industry sponsored R\&D (millions)	&	0.44	&	0.20	&	-0.03	&	0.43	&	0.07	&	1.00	&		&		&		\\
12. Industry contractual ties	&	0.49	&	0.17	&	0.08	&	0.50	&	0.36	&	0.26	&	1.00	&		&		\\
13. NSF grants (log)	&	0.40	&	0.28	&	-0.29	&	0.51	&	-0.16	&	0.25	&	0.22	&	1.00	&		\\
14. NIH grants (log)	&	0.42	&	0.23	&	0.41	&	0.71	&	0.61	&	0.27	&	0.38	&	0.10	&	1.00	\\\hline\hline	
$^\dagger N = 1,165$																				
\end{tabular}																							
}																							
\label{tableDescriptives}																							
\end{sidewaystable}

\begin{table}[h!]
\begin{center}
\begin{tabular}{lccccc}
\multicolumn{6}{c}{Table 4}\\
\multicolumn{6}{c}{Descriptive Statistics for Groups}\\
\multicolumn{6}{c}{Used in Difference-In-Differences (DID) Estimation}\\
\hline \hline
\multicolumn{1}{c}{}&\multicolumn{1}{c}{Mean}&\multicolumn{1}{c}{Median}&\multicolumn{1}{c}{SD}&\multicolumn{1}{c}{Min} &\multicolumn{1}{c}{Max}\\
\hline
Disruptive Patent--Prior Art Pair (``Treated'') &  &	 & 	&& \\
\hspace{4 mm} Disruptive patent grant year & 1994.46 & 1995 &	6.25 & 1977 	&  2006 \\
\hspace{4 mm} Disruptive patent prior art grant year & 1989.24 & 1990	 & 6.55 & 1976	& 2006 \\
\hspace{4 mm} Years between patent and disruptive patent grant & 5.22 &	5 & 3.05	&0&12 \\ 
\hspace{4 mm} Pieces of prior art cited by disruptive patent &4.12  &	3 & 3.61	&1&37 \\ 
\hspace{4 mm} Citations to disruptive patent prior art$_{[t,t-3]}$ & 15.22 &	10 & 16.17	&1&105 \\ 
\hspace{4 mm} Disruptive patent NBER class &  &	 & 	&& \\
\hspace{8 mm} Chemical & 0.23 & 0	 & 0.42 	& 0&1 \\  
\hspace{8 mm} Computers \& communication & 0.09 & 0	 & 0.28 	& 0&1 \\ 
\hspace{8 mm} Drugs \& medical & 0.36 & 0	 & 0.48 	&0&1  \\ 
\hspace{8 mm} Electrical \& electronic & 0.24 & 0	 & 0.43 	& 0&1 \\ 
\hspace{8 mm} Mechanical & 0.04  & 0	 &  0.21	& 0&1 \\ 
\hspace{8 mm} Others & 0.03 & 	0 & 0.18 	& 0&1 \\
\hspace{4 mm} Disruptive patent prior art NBER class &  &	 & 	&& \\
\hspace{8 mm} Chemical & 0.27 & 0	 & 0.45	&0&1 \\  
\hspace{8 mm} Computers \& communication & 0.08 & 0	 & 0.27	&0&1 \\ 
\hspace{8 mm} Drugs \& medical & 0.32 &	0 & 0.47	&0&1 \\ 
\hspace{8 mm} Electrical \& electronic & 0.25 & 	0 & 0.43	&0&1 \\ 
\hspace{8 mm} Mechanical & 0.05 &	0 & 0.21	&0&1 \\ 
\hspace{8 mm} Others & 0.03 & 	0 & 0.02	&0&1 \\
\hline
Control Patent--Prior Art Pair (``Control'') & 	& 	 & && \\
\hspace{4 mm} Control patent grant year & 1994.46 &	1995 & 6.25	&1977&2006 \\
\hspace{4 mm} Control patent prior art grant year  & 1989.24 &	1990 & 6.55	&  1976	& 2006 \\
\hspace{4 mm} Years between patent and control patent grant & 5.22 &	5 & 3.05 	& 0&12 \\ 
\hspace{4 mm} Pieces of prior art cited by control patent & 4.06 &	3 & 3.26	&1&23 \\ 
\hspace{4 mm} Citations to control patent prior art$_{[t,t-3]}$ & 15.21 &	10 & 16.15	&1&108 \\ 
\hspace{4 mm} Control patent NBER class &  &	 & 	&& \\
\hspace{8 mm} Chemical & 0.23 & 0	 & 0.42 	& 0&1 \\ 
\hspace{8 mm} Computers \& communication & 0.09 & 0	 & 0.28 	& 0&1 \\ 
\hspace{8 mm} Drugs \& medical & 0.36 & 0	 & 0.48 	&0&1  \\ 
\hspace{8 mm} Electrical \& electronic  & 0.24 & 0	 & 0.43 	& 0&1 \\
\hspace{8 mm} Mechanical & 0.04  & 0	 &  0.21	& 0&1 \\ 
\hspace{8 mm} Others  & 0.03 & 	0 & 0.18 	& 0&1 \\
\hspace{4 mm} Control patent prior art NBER class &  &	 & 	&& \\
\hspace{8 mm} Chemical & 0.27 & 0	 & 0.45	&0&1 \\
\hspace{8 mm} Computers \& communication  & 0.08 & 0	 & 0.27	&0&1 \\  
\hspace{8 mm} Drugs \& medical & 0.32 &	0 & 0.47	&0&1 \\
\hspace{8 mm} Electrical \& electronic & 0.25 & 	0 & 0.43	&0&1 \\
\hspace{8 mm} Mechanical& 0.05 &	0 & 0.21	&0&1 \\
\hspace{8 mm} Others & 0.03 & 	0 & 0.02	&0&1 \\
\hline\hline

\end{tabular}
\end{center}
\caption{\textsc{Notes.---}Each group consists of 2,746 patent--prior art pairs, for for an effective sample size of size of 5,492.  Disruptive patents are defined here to be those that are one standard deviation above the mean level of disruptiveness among the sample of university patents, conditional on having a positive disruptiveness score and citing at least one piece of prior art.}
\end{table}

\pagebreak
 \clearpage

\newcolumntype{g}{>{\columncolor{Gray}}D{.}{.}{-1}}
\newcolumntype{d}{>{\columncolor{Gray}}c}
\newcolumntype{w}{D{.}{.}{-1}}

\begin{sidewaystable}[h!]\centering
\def\sym#1{\ifmmode^{#1}\else\(^{#1}\)\fi}
\captionsetup{justification=centering}
\caption{Table 5\\ Regression Models of University Patenting, 1993-2005$^\dagger$}
\scalebox{0.68}{
\small\addtolength{\tabcolsep}{-4pt}
\renewcommand{\arraystretch}{1}
\begin{tabular}{lwwwwww}
\hline\hline
               &\multicolumn{1}{c}{Model 1}&\multicolumn{1}{c}{Model 2}&\multicolumn{1}{c}{Model 3}&\multicolumn{1}{c}{Model 4}&\multicolumn{1}{c}{Model 5}&\multicolumn{1}{c}{Model 6}\\
                 &\multicolumn{1}{c}{Negative Binomial}&\multicolumn{1}{c}{Negative Binomial}&\multicolumn{1}{c}{OLS}&\multicolumn{1}{c}{OLS}&\multicolumn{1}{c}{OLS}&\multicolumn{1}{c}{OLS}\\     
                  &\multicolumn{1}{c}{Volume}&\multicolumn{1}{c}{Impact}&\multicolumn{1}{c}{Originality}&\multicolumn{1}{c}{New Combinations}&\multicolumn{1}{c}{Disruptiveness}&\multicolumn{1}{c}{Radicalness}\\       
\hline \\
{\bf Technology transfer experience} &&&&&&\\
[1em]
\hspace{0.2cm}Patent stock$_{t-1}$       &   0.0014\sym{*}  &  -0.0003         &   0.0001         &  -0.0049         &   0.0000         &  -1.2697\sym{***}\\
                & (0.0006)         & (0.0010)         & (0.0003)         & (0.0191)         & (0.0003)         & (0.1498)         \\
[1em]
\hspace{0.2cm}Technology transfer age (decades)$_{(centered)}$           &   0.5694         &   0.2337\sym{***}&   0.0127         &   6.5645\sym{**} &  -0.1304\sym{***}&  61.0185\sym{***}\\
                & (0.4570)         & (0.0589)         & (0.0352)         & (2.0611)         & (0.0324)         &(16.1994)         \\
[1em]
\hspace{0.2cm}Technology transfer age$^2$  (decades)$_{(centered)}$      &  -0.0298\sym{***}&  -0.0225\sym{*}  &  -0.0036         &   0.3851\sym{+}  &   0.0012         & -18.0159\sym{***}\\
                & (0.0062)         & (0.0099)         & (0.0038)         & (0.2252)         & (0.0035)         & (1.7697)         \\
[1em]
\hspace{0.2cm}Life sciences dominant &  -0.0761\sym{**} &  -0.1149\sym{*}  &  -0.0451\sym{***}&   0.5954         &  -0.0038         &  -1.5580         \\
                & (0.0275)         & (0.0484)         & (0.0107)         & (0.6242)         & (0.0098)         & (4.9060)         \\
[1em]
{\bf Scientific capacity} &&&&&&\\
[1em]
\hspace{0.2cm}Scientific articles$_{t-1}$ (log)   &   0.0016         &   0.5075\sym{***}&   0.0012         &  -1.9103         &   0.0398         &  -7.5977         \\
                & (0.1759)         & (0.1268)         & (0.0560)         & (3.2813)         & (0.0516)         &(25.7895)         \\
[1em]
\hspace{0.2cm}Impact factor   &   0.0051\sym{+}  &   0.0303\sym{***}&  -0.0026\sym{*}  &  -0.0919         &   0.0031\sym{**} &   5.0206\sym{***}\\
                & (0.0027)         & (0.0046)         & (0.0010)         & (0.0599)         & (0.0009)         & (0.4710)         \\
[1em]
{\bf Industry ties} &&&&&&\\
[1em]
\hspace{0.2cm}Industry sponsored R\&D (millions) & 0.0013         &   0.0038\sym{*}  &  -0.0004         &   0.0512\sym{+}  &  -0.0000         &  -0.9919\sym{***}\\
                & (0.0011)         & (0.0019)         & (0.0005)         & (0.0300)         & (0.0005)         & (0.2360)         \\
[1em]
\hspace{0.2cm}Industry contractual ties&  -0.0093\sym{*}  &  -0.0157\sym{*}  &  -0.0005         &  -0.2666\sym{+}  &  -0.0001         &  -6.7401\sym{***}\\
                & (0.0046)         & (0.0079)         & (0.0025)         & (0.1478)         & (0.0023)         & (1.1615)         \\
[1em]
{\bf Government ties} &&&&&&\\
[1em]
\hspace{0.2cm}NSF grants (log)  &   0.0466         &  -0.0272         &   0.0264\sym{*}  &   0.0807         &  -0.0039         &  11.0251\sym{*}  \\
                & (0.0299)         & (0.0362)         & (0.0109)         & (0.6408)         & (0.0101)         & (5.0366)         \\
[1em]
\hspace{0.2cm}NIH grants (log)  &   0.0874         &  -0.0105         &   0.0069         &  -0.9679         &   0.0346\sym{+}  &  19.4857\sym{*}  \\
                & (0.0595)         & (0.0568)         & (0.0198)         & (1.1588)         & (0.0182)         & (9.1080)         \\
[1em]
\hspace{0.2cm}Government interest patents&   0.0157\sym{***}&   0.0143\sym{***}&  -0.0000         &  -0.0095         &   0.0002         &   0.6960\sym{**} \\
                & (0.0009)         & (0.0015)         & (0.0005)         & (0.0307)         & (0.0005)         & (0.2410)         \\
[1em]
{\bf Fixed effects} &&&&&&\\
[1em]																									
\hspace{0.2cm}University	&	\multicolumn{1}{c}{\text{Yes}}		&	\multicolumn{1}{c}{\text{Yes}}		&	\multicolumn{1}{c}{\text{Yes}}		&	\multicolumn{1}{c}{\text{Yes}}		&	\multicolumn{1}{c}{\text{Yes}}		&	\multicolumn{1}{c}{\text{Yes}}		\\
[1em]																									
\hspace{0.2cm}Year 	&	\multicolumn{1}{c}{\text{Yes}}		&	\multicolumn{1}{c}{\text{Yes}}		&	\multicolumn{1}{c}{\text{Yes}}		&	\multicolumn{1}{c}{\text{Yes}}		&	\multicolumn{1}{c}{\text{Yes}}		&	\multicolumn{1}{c}{\text{Yes}}	\\
[1em]
Constant        &   2.0468\sym{**} &  -0.8332         &   0.3726\sym{+}  &  16.3437         &  -0.3022         &-205.9845\sym{*}  \\
                & (0.6682)         & (0.6206)         & (0.2063)         &(12.0872)         & (0.1902)         &(95.0010)         \\
\hline
\(N\)    &     1165         &     1165         &     1165         &     1165         &     1165         &     1165         \\
Log likelihood              &-3227.9247         &-5845.0454         &1128.1306         &-3613.8733         &1222.8103         &-6015.8026         \\
\(R^2\)             &                  &                  &   0.1947         &   0.1794         &   0.3503         &   0.5267         \\
\hline\hline 
\multicolumn{7}{l}{\footnotesize \sym{+} \(p<0.1\), \sym{*} \(p<0.05\), \sym{**} \(p<0.01\), \sym{***} \(p<0.001\); two tailed tests.}\\
\multicolumn{7}{p{15cm}}{$^{\dagger}$ Standard errors are in parentheses.}\\
\end{tabular}

}

\end{sidewaystable}

\end{document}